\documentclass{article}

\usepackage{arxiv}
\usepackage[utf8]{inputenc} 
\usepackage[T1]{fontenc}    
\usepackage{hyperref}       
\usepackage{url}            
\usepackage{booktabs}       
\usepackage{amsfonts}       
\usepackage{nicefrac}       
\usepackage{microtype}      
\usepackage{lipsum}		
\usepackage{graphicx}
\usepackage{natbib}
\usepackage{amssymb,amsmath}

\usepackage{caption}

\usepackage{graphicx,times}
\usepackage{natbib}

\usepackage{cancel}

\usepackage{xcolor}
\usepackage{ulem}

\graphicspath{{./}{figures/}}

\usepackage{indentfirst}
\title{The impact of tidal migration of hot Jupiters on the rotation of Sun-like main-sequence stars}

\author{{\includegraphics[scale=0.06]{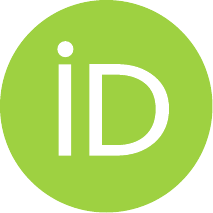}\hspace{1mm}Shuai-Shuai Guo}\thanks{Use footnote for providing further
		information about author (webpage, alternative
		address)---\emph{not} for acknowledging funding agencies.} \\
	Yunnan Observatories, Chinese Academy of Sciences, P.O. Box 110, Kunming 650011, People's Republic of China\\
	School of Astronomy and Space Science, University of Chinese Academy of Sciences, Beijing, People's Republic of China\\
	Key Laboratory for the Structure and Evolution of Celestial Objects, CAS, Kunming 650011, People's Republic of China \\
	\texttt{guoshuaishuai@ynao.ac.cn} \\
}

\renewcommand{\shorttitle}{\textit{arXiv} Template}

\hypersetup{
pdftitle={A template for the arxiv style},
pdfsubject={q-bio.NC, q-bio.QM},
pdfauthor={David S.~Hippocampus, Elias D.~Striatum},
pdfkeywords={First keyword, Second keyword, More},
}
\pdfoutput=1
\begin{document}
\maketitle

\begin{abstract}
	The tidal interactions of planets affect the stellar evolutionary status and the constraint of their physical parameters by gyrochronology. In this work, we incorporate the tidal interaction and magnetic braking of the stellar wind into MESA and calculate a large grid of 25000 models, covering planets with masses of 0.1-13.0$\,$$M_{\mathrm{J}}$ with different orbital distances that orbit late-type stars of different metallicities. We also explore the effect of different stellar initial rotations on the tidal interactions. Our results show that in the case of tidal inward migration, the stellar rotation periods are always lower than that of the star without planet before the planet is engulfed and the difference in the rotation period of its host star always increases with time. After the planet is engulfed, the stellar rotation periods are still lower than that of star without planet, but the difference of periods can be quickly eliminated if the star has a thick convective envelope(smaller mass and larger metallicity), regardless of the mass of the planet and the initial rotation period of the star. In the case of stars with thinner convective envelopes(larger mass and smaller metallicity), the stars will be spined up and remain the faster rotation in a long time. Meanwhile, the planet is easily swallowed and the period differences are large if the initial rotation period of its host star is higher. Final, we also study the evolution of WASP-19 and estimate the range of tidal quality parameter $Q'_{*} = (4.6 \pm 0.9) \times 10^{6}$ and initial semi-major axis as $(0.035 \pm 0.004)$$\,$au.
\end{abstract}


\section{Introduction}
Up to now, more than 5000 exoplanets have been confirmed since the discovery of the first exoplanet 51Pegasi b \citep{1995Natur.378..355M}. The Kepler, CoRoT, TESS and ground-based telescopes have enabled astronomers to have a deeper insight into the exoplanet systems. A large number of close-in hot Jupiters (P $<$ 10$\,$days) around the low-mass stars have been found \citep{2012Natur.486..375B,2013ApJ...766...81F,2018AJ....156..254W,2019AJ....158..141Z}. According to the current theory of planet formation, it is difficult to accumulate materials required for planet formation due to such a short distance. These planets are unlikely to form at the current location \citep{1996Natur.380..606L}. Thus, in the early phase of planet formation, hot Jupiter are thought to be migrated to about 0.1$\,$au due to the drag of the gas nebula in the protoplanetary disk \citep{1996Natur.380..606L,1998ApJ...500..428T,1997Icar..126..261W,2007A&A...463..775P}. After the dissipation of protoplanetary disk, the tidal interactions between close-in planets and their host stars become important, which results in two important consequences. First, the eccentricities of the hot Jupiters tend to be low \citep{2015ARA&A..53..409W}, and the interactions between star and planet also change the orbital semi-major aixs $a$ \citep{2008ApJ...678.1396J,2009ApJ...692L...9L,2010ApJ...712.1107G}. Second, growing evidences suggest that hot Jupiters play an important role in modifying the stellar rotation. For example, \citet{2010MNRAS.408.1770A} first discovered that the rotations of stars with planets are faster than that without planets. Recently, \citet{2021ApJ...919..138T} also found that the host stars of hot Jupiters have faster rotations by analysing the results of the California Kepler Survey (CKS; \citep{2017AJ....154..107P}) and the SWEET-Cat catalog \citep{2013A&A...556A.150S}.

The rotation of stars has been studied for more than a century, and such study has been rapidly developed in recent decades. For stars with masses of 0.3$\,$$M_{\odot}$ $\leq$ M $\leq$$\,$1.3$\,$$M_{\odot}$, they are consisted of a central  radiative core and a convective envelope. When the mass of the star is less than 0.3$\,$$M_{\odot}$, the star is fully convective. The convective envelopes of these low-mass stars contain surface magnetic fields that interact with the stellar wind to extract angular momentum from the stars. As a result of the loss of angular momentum, the rotation of the star decreases with time. This is often called magnetic braking \citep{1967ApJ...150..551K,1972ApJ...171..565S,1988ApJ...333..236K}. Many models have been developed \citep{2015ApJ...799L..23M,2013ApJ...776...67V,2018ApJ...862...90G}, for instance, \citet{2020ApJ...889..108A} discussed the evolution of the rotation period of Solar-like stars with different metallicities, and found that metal-poor stars tend to rotate faster. As an application of magnetic braking, one can determine the age of the star through the evolutionary characteristics of its rotation period because the stellar rotation period will converge to a certain value with the evolution owing to the magnetic braking. This method is named gyrochronology \citep{2003ApJ...586..464B,2007ApJ...669.1167B,2010ApJ...722..222B,2008ApJ...687.1264M}. The accuracy of the age obtained by this method is strongly dependent on the completeness of the stellar model. Although the rotation periods of some open clusters have been measured ( Praesepe: 670$\,$Myr \citep{2017ApJ...842...83D}; the Hyades: 30$\,$Myr \citep{2019ApJ...879..100D}; NGC 6811: 1.0$\,$Gyr \citet{2019ApJ...879...49C} and NGC 752: 1.4$\,$Gyr \citep{2018ApJ...862...33A} ), \citet{2019ApJ...879...49C} found that for low-mass stars the magnetic braking models tend to predict too much angular momentum loss so that the inferred ages are likely to be incorrect. In fact, the orbital decay of planet can rise the rotation of host star \citep{2007P&SS...55..643C}. \citet{2014MNRAS.442.1844B} compared the ages of the host stars by using the isochrone and gyrochronology on a sample of 68 planet-hosting stars and found that the ages obtained by gyrochronology are smaller than that obtained by isochrone, which can be an indirect evidence that star-planet interaction speeds up the stellar rotation. Furthermore, by including the tidal interaction, \citet{2019A&A...626A.120G} and \citet{2020A&A...641A..38G}corrected the ages of gyrochronology for some stars with planet, which provides a standalone tool based on tidal-chronology to estimate the age of a massive close-in planetary system. In addition, the modification of the stellar rotation produced by the planetary tides is also related to the metallicity of the star. \citet{2017A&A...604A.113B} and \citet{2020A&A...643A..34O} show that for 1.0$\,$$M_{\odot}$ stars, planetary engulfment events are more likely to occur on metal-poor stars. Thus, the faster rotation of metal-poor stars can be attributed to the increase of the stellar angular momentum due to planetary engulfment. In fact, the interaction between the planet and its host star is affected by many factors. Therefore, our aim is to study the influence of those factors on the stellar rotation. We calculate a large sample, in which we explore the influences of free parameters ($M_{*}$, $[Fe/H]$, $M_{\mathrm{pl}}$, $a_{\mathrm{ini}}$, $Q'_{*}$) on the rotation of host star. In addition, we also check the effect of stellar tides on the rotation before and after engulfment for different star-planet systems in details.

In this paper, the structure is as follows. In Section \ref{sec:interaction}, we describe the model of star-planet interaction. In Section \ref{sec:metallicity}, we have discussed how metallicity influences the evolution of the interaction model. In Section \ref{sec:age}, we discussed the influences of 6 parameters on the variation of rotation periods of star with age and the evolution of WASP-19. In Section \ref{sec:discussion}, we discussed the limitations of our model. In Section \ref{sec:conclusion}, we make conclusions about this work.


\section{Star-planet interaction model in MESA} \label{sec:interaction}
In this section, we study the exchange of angular momentum between the planetary orbit and the host star's rotation. The angular momentum of planet itself is neglected because it is small compared to that of the former. In fact, the timescale of the exchange of angular momentum is comparable with the evolutionary timescale of the star. We thus implement the processes of the exchange of angular momentum into the stellar evolution code MESA, version 11554. So far, MESA has been widely used in the field of stellar and planetary evolution \citep{ 2011ApJS..192....3P,2013ApJS..208....4P,2015ApJS..220...15P,2018ApJS..234...34P,2019ApJS..243...10P}. For the stellar evolution, we consider the phase of main sequence for the stars with masses of 0.8-1.3$\,$$M_{\odot}$. We adopt the protosolar abundances of \citet{2009ARA&A..47..481A} as a reference for all metallicities, and here we use Table 1 of \citet{2020ApJ...889..108A} as the input for the metallicity in our model. The stellar metallicity [Fe/H] is in the range of  -0.5 to +0.5$\,$dex. For stellar element abundances, we use solar metallicities Z = 0.0134 as the benchmark. We set 0.6-16 days as the initial rotation periods of the star. In order to simulate the change of the periods of the star and planetary orbit caused by the tidal friction, we only consider massive planets with a mass range of 0.1-13.0$\,$$M_{J}$. For planets with small orbital distances (a \textless 0.1$\,$au), the influence of tidal force is prominent \citep{2021ApJ...919..138T}. Therefore, the initial orbital distance is set to be 0.01-0.10$\,$au. In calculating the tidal dissipation rate \citep{1963MNRAS.126..257G}, the tidal quality parameter $Q'_{*}$ of the star is a very important parameter. A smaller $Q'_{*}$ value means a larger tidal dissipation rate, and vice versa. However, $Q'_{*}$ is still an indeterminate quantity from both theoretical and observational sides. \citet{2020AJ....159..150P} used the transit-timing method to detect the tidal quality parameter $Q'_{*}$ for 12 close-in hot Jupiter systems. Their results show that the values of $Q'_{*}$ are larger than 10$^{5}$. \citet{2019AJ....158..190H} constrain the value of $Q'_{*}$ empirically and suggest that $Q'_{*}$ varies from $10^{5}$ to $10^{9}$. Thus, we also explore a same range of the tidal quality parameter $Q'_{*}$ ($10^{5}$ - $10^{9}$). We summarize the values that we have adopted in Table \ref{tab:tab1}.

\subsection{Stellar rotation and wind braking model} \label{subsec:braking}
We implement the magnetic braking model of \citet{2015ApJ...799L..23M} into the rotation scheme provided in MESA. We simulate the entire evolution of stars from pre-main sequence(PMS) to end of the main sequence(MS). In the PMS phase, the angular momentum is transferred between the stars and their protostellar disks and the star maintains a constant rotation rate in the disk locking time. In this paper, the disk locking time scale for all models is 13$\,$Myr \citep{2021ApJ...912...65G}. During this period, there is no exchange of angular momentum with the gas around the star. Thus, the stellar angular momentum is assumed to be constant. After the disk is dissipated, the magnetic braking results in the loss of the angular momentum in stars. We use magnetic braking model of \citet{2015ApJ...799L..23M} scheme as following:

\begin{equation}
\frac{\mathrm{d}L_{wind}}{\mathrm{d}t}=-T_0(\frac{\tau_{\mathrm{cz}}}{\tau_{\mathrm{cz\odot}}})^p(\frac{\Omega_*}{\Omega_\odot})^{p+1}		(unsaturated),
\label{eq1}
\end{equation}

\begin{equation}
\frac{\mathrm{d}L_{\mathrm{wind}}}{\mathrm{d}t}=-T_0{\chi}^p(\frac{\Omega_*}{\Omega_\odot})		(saturated),
\label{eq2}
\end{equation}

\begin{equation}
T_0=K(\frac{R_{*}}{R_\odot})^{3.1}(\frac{M_{*}}{M_\odot})^{0.5}{\gamma}^{2m},
\label{eq3}
\end{equation}

\begin{equation}
\gamma=\sqrt{1+(u/0.072)^2},
\label{eq4}
\end{equation}
\begin{equation}
u=\frac{\Omega_{*}}{\Omega_{\mathrm{crit}}},\Omega_{\mathrm{crit}}=\sqrt{\frac{GM_*}{{R_*}^3}},
\label{eq5}
\end{equation}
where $L_{\mathrm{wind}}$ is the angular momentum lost by the magnetic stellar wind, $R_*$, $\Omega_{*}$, and $\Omega_{\mathrm{crit}}$ are the stellar radius, angular rotation rate, and critical angular rotation rate, respectively. The constants $K$, $m$, $p$, and $\chi$ are free parameters. The saturated and unsaturated states in Equations (\ref{eq1}) and (\ref{eq2}) represent two different states of stellar magnetic activity, both of which have a strong correlation with the Rossby number $R_{\mathrm{o}}=\frac{2\pi}{\Omega_{\mathrm{crit}}\tau_{\mathrm{cz}}}$. The convective turnover timescale $\tau_{\mathrm{cz}}$ can be expressed as \citep{2021ApJ...912...65G}:
\begin{equation}
\tau_{\mathrm{cz}}(r)=\alpha_{\mathrm{MLT}}H_{\mathrm{P}}(r)/v_{\mathrm{c}}(r),
\label{eq6}
\end{equation}
where $H_{\mathrm{P}}(r)$ is the scale height, $v_{\mathrm{c}}(r)$ is the convective velocity at radius $r$, and $\alpha_{\mathrm{MLT}}$ is the convective mixing length. In Equation (\ref{eq6}), $\tau_{\mathrm{cz}}$ is the turnover timescale of the convective zone, which is defined in the location where $r$ = $r_{\mathrm{BCZ}}$ + $0.5H_{\mathrm{P}}(r)$, and $r_{\mathrm{BCZ}}$ is the radius of the bottom of the outer convection zone. The values of $\alpha_{\mathrm{MLT}}$ is 1.82.

$R_{\mathrm{o}}$ $\leq$ $R_{\mathrm{osat}}$ is the saturation region, and vice versa. The parameter $\chi$ = $R_{\mathrm{o}}$ /$R_{\mathrm{osat}}$ relies on the critical value $R_{\mathrm{osat}} = 0.14$ \citep{2018MNRAS.479.2351W}. The solar $R_{\mathrm{o}}$ is around 2 \citep{2016MNRAS.462.4442S}. Thus we take $\chi$ = 14. The constant $m$ is set to be 0.22. In order to reproduce the current rotation period of the Sun, the values of $K$ and $p$ are 1.2 $\times$ $10^{30}$ and 2.6, respectively.

\subsection{Orbital evolution and angular momentum transfer} \label{subsec:orbital}
For the sake of simplicity, we only consider the circular orbits, i.e. the eccentricity is zero. We assume that the star is a spherically symmetrical and uniformly rotating object, and the planet is a mass point. For an isolated star-planet system, the total angular momentum is not conservative and can be reduced by the magnetic wind. We have the following expression:
\begin{equation}
\frac{\mathrm{d}L_{*}}{\mathrm{d}t}=\frac{\mathrm{d}L_{\mathrm{wind}}}{\mathrm{d}t}-\frac{\mathrm{d}L_{\mathrm{orb}}}{\mathrm{d}t}-\frac{\mathrm{d}L_{\mathrm{sp}}}{\mathrm{d}t},
\label{eq7}
\end{equation}

\begin{equation}
L_{*}=I_*\Omega_{*},
\label{eq8}
\end{equation}
\begin{equation}
L_{\mathrm{sp}}=I_{\mathrm{sp}}\Omega_{\mathrm{sp}},
\label{eq9}
\end{equation}
\begin{equation}
L_{\mathrm{orb}}=\frac{M_{*}M_{\mathrm{pl}}}{M_{*}+M_{\mathrm{pl}}}na^2,
\label{eq10}
\end{equation}
\begin{equation}
n=\sqrt{\frac{G(M_*+M_{\mathrm{pl}})}{{a}^3}},
\label{eq11}
\end{equation}
where $L_{*}$, $L_{\mathrm{sp}}$ and $L_{\mathrm{orb}}$ are the stellar, planetary and orbital angular momentum, respectively. $I_*$ and $I_{\mathrm{sp}}$ are the stellar and planetary moment of inertia. $n$ is the orbital angular frequency. $a$ is the semi-major axis, and $M_{\mathrm{pl}}$ is the planetary mass. Since the rotational angular momentum of the planet is negligible compared to its orbital angular momentum, we assume $L_{\mathrm{sp}}$ = 0. Combining Equations (\ref{eq7})-(\ref{eq11}), we obtain the expression of the time derivative of the
stellar angular rotation rate:
\begin{equation}
\frac{\mathrm{d}\Omega_{*}}{\mathrm{d}t}=\frac{1}{I_*}(\frac{\mathrm{d}L_{\mathrm{wind}}}{\mathrm{d}t}-\Omega_{*}\frac{\mathrm{d}{I_*}}{\mathrm{d}t}-\frac{1}{2}M_{\mathrm{pl}}M_*\sqrt{\frac{G}{a(M_*+M_{\mathrm{pl}})}}\frac{\mathrm{d}a}{\mathrm{d}t}).
\label{eq12}
\end{equation}
From Equation (1) of \citet{2012ApJ...751...96P}, we obtain the evolution of the semi-major axis $a$ \citep{1963MNRAS.126..257G,1968aitp.book.....K,2008ApJ...678.1396J}:
\begin{equation}
\frac{\mathrm{d}a}{\mathrm{d}t} = sign(\Omega_{*}-n) \frac{9}{2} \sqrt{\frac{G}{aM_*}} (\frac{R_*}{a})^5 \frac{M_{\mathrm{pl}}}{Q'_*},
\label{eq13}
\end{equation}
where $sign(\Omega_{*}-n)$ takes the value of 1 when the star spins faster than the planet and -1 when the reverse is true. We use the method of \citet{2020A&A...643A..34O} to set the Roche limit at 1.44 times the radius of the star, that is, when the semi-major axis $a$ = 1.44$\,$$R_*$, the planet is destroyed. However, it should be noted that we adopt the Roche limit as 1.44 times the stellar radius, consistent with the approach of \citet{2020A&A...643A..34O}. When the planet reaches the Roche limit, the transfer of orbital angular momentum to the star stops, and the remaining orbital angular momentum is removed from the system. At this point, the rotational velocity of the star when the planet is consumed represents a lower limit. 

\begin{table}
\begin{center}
\caption[]{ The parameters adopted in this work}\label{tab:tab1}
 \begin{tabular}{clclclcl}
  \hline\noalign{\smallskip}
Parameter &  &  &  &  &  &  &                    \\
  \hline\noalign{\smallskip}
$M_{*}$$\,$($M_{\odot}$) & 0.8 & 0.9 & 1.0 & 1.1 & 1.2 & 1.3 & --  \\
$[Fe/H]$$\,$(dex) & -0.5 & -0.3 & 0.0  & +0.3 & +0.5 & -- & -- \\
$M_{\mathrm{p}}$$\,$($M_{\mathrm{J}}$)& 0.1 & 0.5 & 1.0 & 4.0 & 8.0 & 13.0 & --   \\
$P_{\mathrm{rot, ini}}$$\,$(days) & 0.6 & 3.0 & 8.0 & 16.0 & -- & -- & --   \\
$Q_{*}$ & $10^{5}$ & $10^{6}$ & $10^{7}$ & $10^{8}$ & $10^{9}$ &-- &--  \\
$a_{\mathrm{ini}}$$\,$(au) & 0.01 & 0.02 & 0.03 & 0.04 & 0.05 & 0.06 & 0.10   \\
  \noalign{\smallskip}\hline
\end{tabular}
\end{center}
\end{table}

\section{How metallicity affect the evolution of star-planet systems} \label{sec:metallicity}

\begin{figure}
  \begin{minipage}[t]{\linewidth}
  \centering
\captionsetup{labelsep=none}
   \caption*{(a)}
   \includegraphics[width=0.7\linewidth]{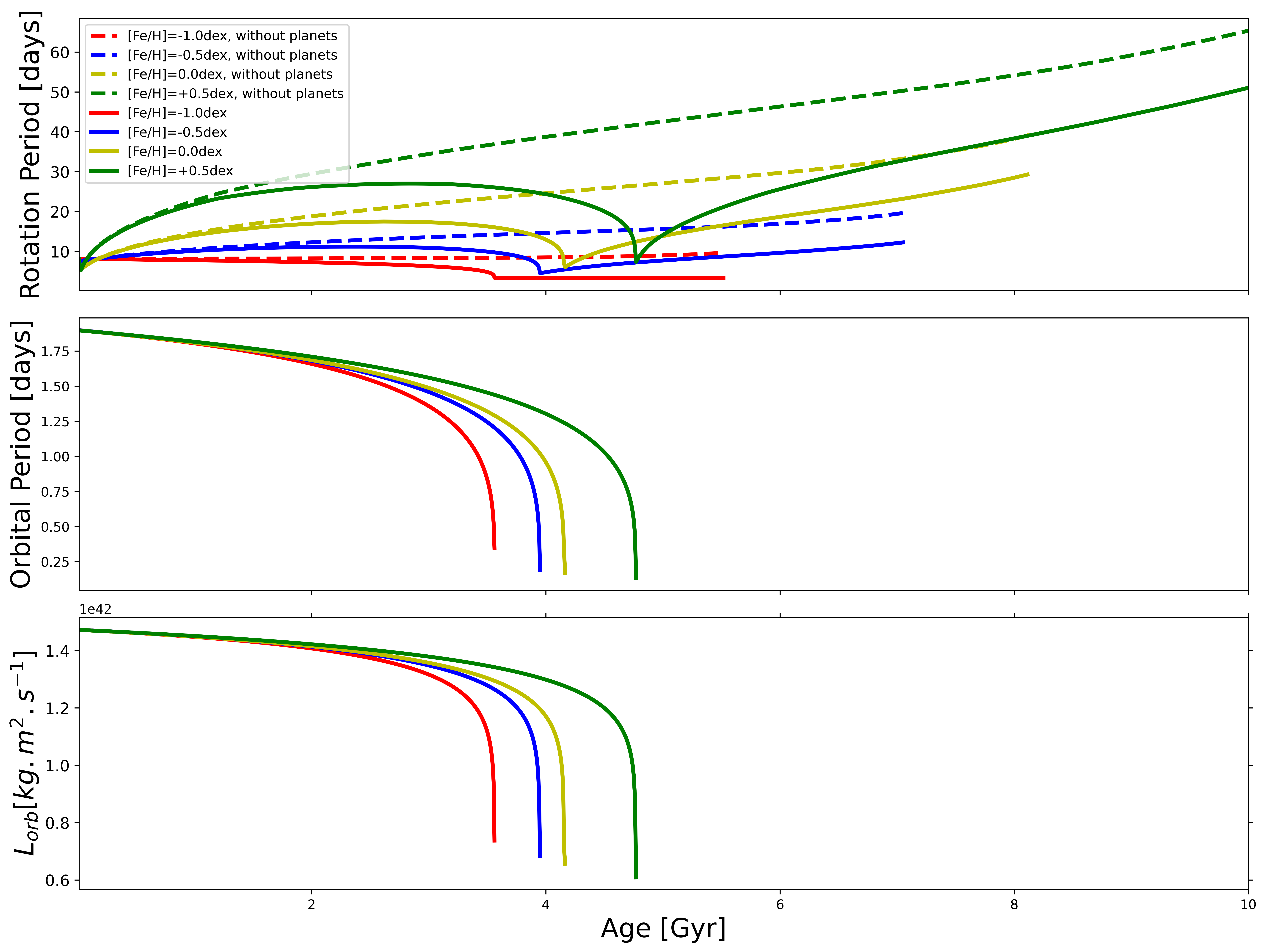}
  \end{minipage}%
	
\begin{minipage}[t]{\linewidth}
  \centering
\captionsetup{labelsep=none}
   \caption*{(b)}
   \includegraphics[width=0.7\linewidth]{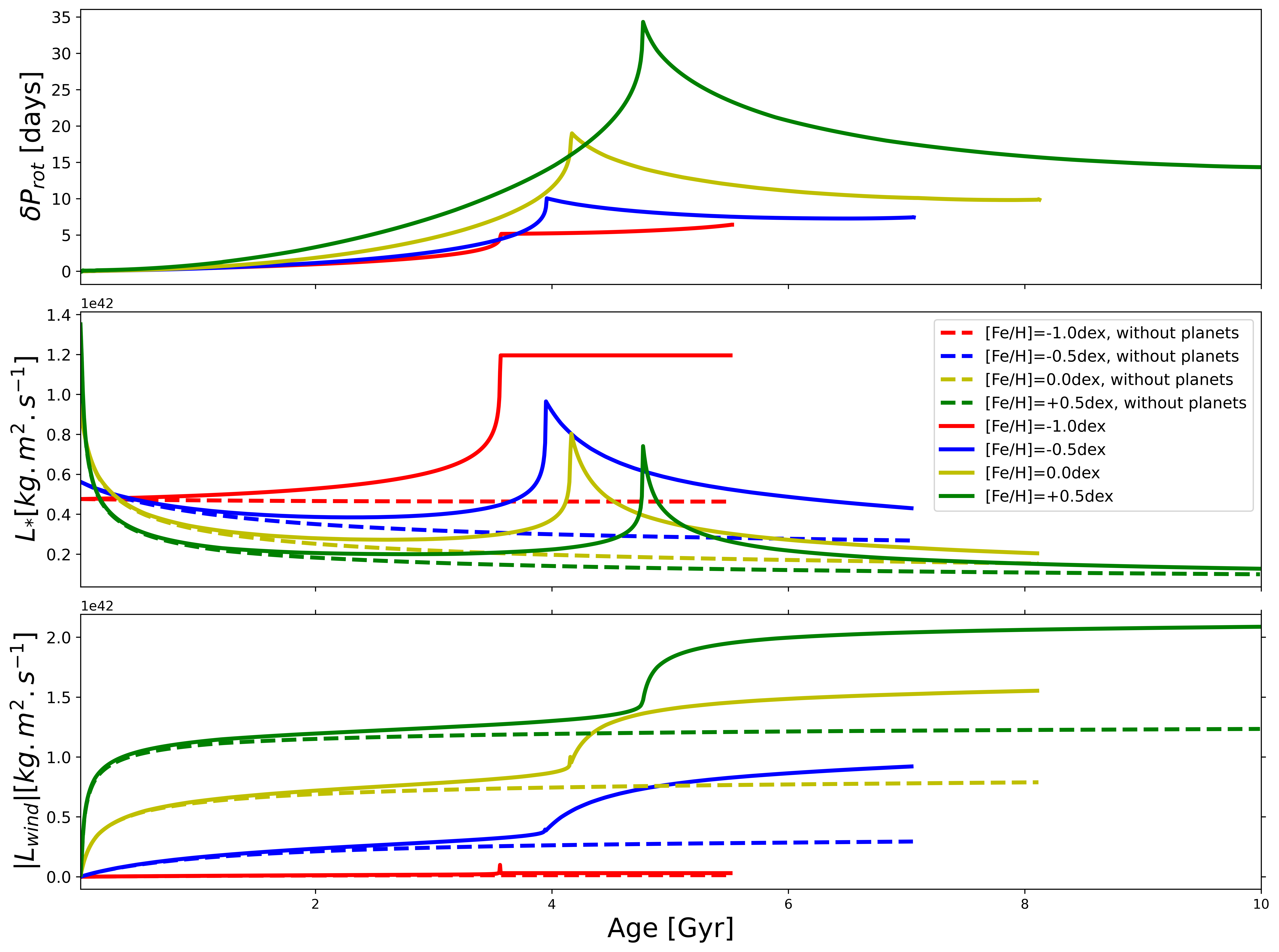}
  \end{minipage}%
\caption{In panel (a), the top panel shows the relationship between the evolution time and rotation period for individual stars and star-planet systems. The middle panel shows the relationship between the evolution time and planetary orbital period. The bottom panel shows the relationship between the evolution time and orbital angular momentum $L_{\mathrm{orb}}$. Solid and dashed lines represent stars with and without planets at different metallicities, respectively. Here, the mass of the star is 1.0$\,$$M_{\odot}$, and the metallicity is represented by the red, blue, yellow, and green lines for [Fe/H] = -1.0, -0.5, 0.0, and +0.5$\,$dex, respectively. The planetary mass is 1.0$\,$$M_{\mathrm{J}}$, the initial semi-major axis is 0.03$\,$au, the initial rotation period of the star is 8.0 days, and the tidal quality parameter $log_{{10}}\mathcal{Q'}{_*}$ is 7.
In panel (b), the top panel shows the evolution of $\delta P$ as a function of time. The middle panel shows the relationship between the evolution time and stellar spin angular momentum $L_*$. The bottom panel shows the relationship between the evolution time and magnetic braking angular momentum loss $\lvert L_{\mathrm{wind}} \rvert$. Here, $\delta P$ represents the difference in rotation periods of stars with and without planets, quantifying the influence of the planet on stellar rotation.
\label{fig:delta}}
\end{figure}

In this section, we have extensively discussed the impact of different metallicities on the tidal interactions between solar-mass stars and close-in hot Jupiters. This analysis aims to provide a comprehensive understanding of how various physical quantities evolve in our model. Figure \ref{fig:delta} (a) top depicts the influence of star-planet interactions on stellar rotation for stars with different metallicities. These stars without planets show the increasing trend for their rotation periods due to the magnetic braking. In fact, the metallicity dramatically influences the spin down of the stars. The spin down of metal-rich stars are much faster than that of metal-poor stars because of their deeper convective zones. The evolutionary path of the rotation period is affected when a planet with the mass of Jupiter is included. For metal-rich stars with and without planets, the divergence of periods occurs at about 1.0$\,$Gyr. In the case, the orbital decay of the close-in planets transfers angular momentum to the stars so that the rotation of stars are accelerated until the planets spiral down to the stars. Subsequently, the periods of the stars increase again. In contrast, metal-poor stars with solar masses have a larger radius and stronger tidal influences (Equation \ref{eq13}), which are more likely to result in an early planetary engulfment, in the medium panel of Figure \ref{fig:delta} (a), stars with lower metallicities have faster planetary orbit decay and are more likely to experience early planetary engulfment, as predicted by \citet{2020A&A...643A..34O}. In the bottom panel of Figure \ref{fig:delta} (a),  as planets migrate inward, their orbital angular momentum is transferred to the star and gradually decreases over time. Additionally, for larger radius stars, their Roche limit is further from the star, so the transfer of orbital angular momentum to the star is slightly less for larger radius stars. We also note that the modification of the stellar rotation period by the planet is reduced with the decrease of the metallicity.
The difference of the rotation periods of the stars with and without planets, $\delta P$, is given by:
\begin{equation}
\delta P = P_{\mathrm{without}} - P_{\mathrm{with}},
\label{eq14}
\end{equation}
where $P_{\mathrm{without}}$ and $P_{\mathrm{with}}$ are the rotation periods of the stars without and with a planet. Figure \ref{fig:delta} (b) top shows the difference of the rotation period. One can see that the  planets  have more effect on the rotation of metal-rich stars than that of metal-poor stars. This is because the metal-poor stars have thinner convective envelopes, which results in much weaker magnetic braking and thus week effects of the planetary angular momentum transfer on the stellar rotation. In the medium and bottom panels of Figure \ref{fig:delta} (b), we can see that for stars with planets, the rotational angular momentum decreases rapidly for all stars except those with [Fe/H] = -1.0$\,$dex within 2.0$\,$Gyr. This phenomenon is mainly due to two reasons: firstly, the initial rotation speed of the star is faster, resulting in stronger magnetic braking and more severe loss of angular momentum; secondly, in the early stages, the planet is far from the star, and the tidal force is weak. Thus, the angular momentum transferred to the star is smaller than the angular momentum lost due to magnetic braking, resulting in a slower rotation of the star. As the evolution progresses, the wind torque decreases while the tidal torque increases until the tidal torque exceeds the wind torque at around 2.0$\,$Gyr, causing an increase in the star's rotation. After the planet is engulfed, only magnetic braking can cause angular momentum loss, and tidal torque disappears. The star with [Fe/H] = -1.0$\,$dex has a very thin convective envelope, which results in almost no loss of angular momentum. This system can be approximately treated as conserving angular momentum, whereby the stellar angular momentum increases due to the transfer of orbital angular momentum before the planet is engulfed. After the planet is engulfed, the stellar angular momentum remains constant.

\section{The variation of stellar rotation periods of star with age} \label{sec:age}
Our goal is to investigate the influence of tidal interaction on the rotation of star and orbit of planet at different metallicities. Thus we compare the evolution of rotation period of the star for the cases with and without a planet.
In the above model, there are six free parameters, namely, the initial rotation period of the star, the initial orbital distance, the masses of the star and planet, the metallicity, and tidal quality parameter $Q'_{*}$.

In this section, we will discuss how the stellar rotations are affected by the tidal interaction. We define the difference of age $\delta Age$ as:
\begin{equation}
\delta Age = Age_{\mathrm{now}} - Age_{\mathrm{planet-engulf}},
\label{eq15}
\end{equation}
where $Age_{\mathrm{now}}$ is the current age of the star. We choose 0.5, 1.0, 2.0, 3.0, 4.0, 5.0, 6.0, 7.0, 8.0, 9.0 and 10.0$\,$Gyr from the evolutionary models as the comparison samples. $Age_{\mathrm{planet-engulf}}$ is the time when the planet is engulfed. For each age interval, $\delta$Age = 0 is the moment that the planet has just been engulfed. At the same time, all angular momentum of planet is transferred to the host star. In addition, the stellar $\delta$P reached the maximum value in the entire evolution stage (see Figure \ref{fig:delta} (b) top) at this moment.  The differences of stellar periods are showed in Figures \ref{fig:1.0} and \ref{fig:0913}. The red dashed lines in each panel denote the engulfment moment, and the left and right sides of the lines are in the evolution stage before and after the planetary engulfment.

\subsection{The effects of initial semimajor axis} \label{subsec:axis}

\begin{figure}
\centering
\includegraphics[width=\textwidth, angle=0]{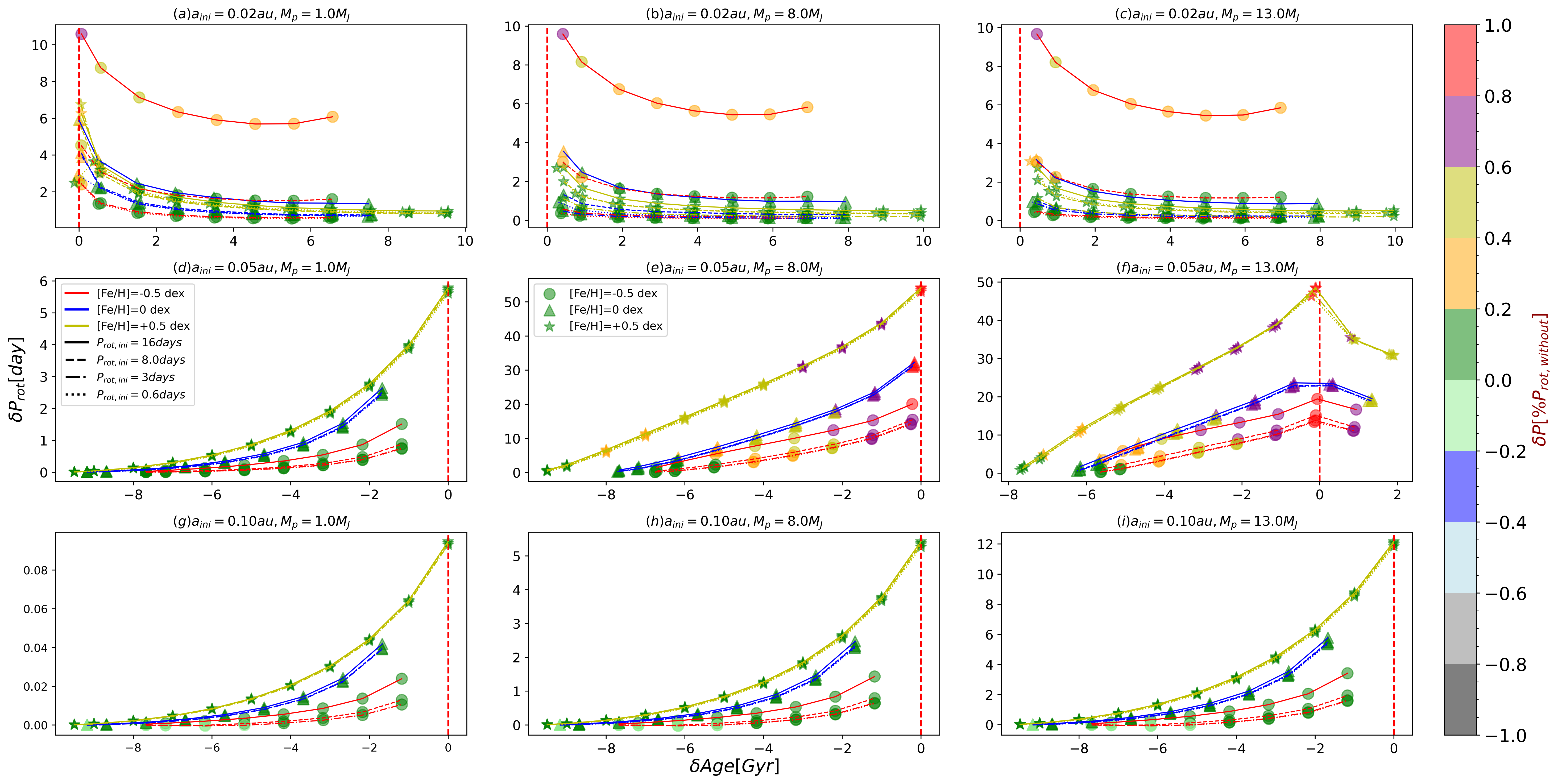}
\caption{The relationship between $\delta Age$ and $\delta P$ of a star with a mass of 1.0$\,$$M_{\odot}$, tidal quality parameter $log_{_{10}}\mathcal{Q'}_{_*}$ = 7, initial semi-major axis of 0.02, 0.05, 0.10$\,$au, respectively. and planetary mass of 1.0, 8.0, 13.0$\,$$M_{\mathrm{J}}$, respectively.  The red dashed lines denote the time of engulfment. The left and right points of the red dashed line means that the planets are alive or engulfed. The circles, triangles and stars represent the stars with metallicity [Fe/H] = -0.5, 0, +0.5$\,$dex, respectively; the solid lines, dashed lines, dot-dashed lines, and dotted lines respectively represent the initial rotation period of the stars of 16.0, 8.0, 3.0, 0.6$\,$days. Colorbar represents the relative rate of change of stellar rotation period, i.e. $\delta P[\% P_{\mathrm{rot,without}}] = \delta P/P_{\mathrm{without}}$.
\label{fig:1.0}}
\end{figure}

First, it can be seen from  Figure \ref{fig:1.0} that for the initial semi-major axis of 0.02$\,$au, a planet with a mass larger than 1.0$\,$M$_{J}$ can be engulfed before 0.5$\,$Gyr. Interestingly, $\delta P$ is almost independent of the mass of the planet. After the engulfment ($\delta$ Age = 0 denotes the time of engulfment.), $\delta P$ decreases quickly with the evolution owing to the magnetic braking. In fact, the process of the magnetic braking is proportional to the angular frequency of the stellar rotation. When the planet is swallowed up, the angular frequency is enhanced rapidly.  As a consequence, the stellar rotation period drops faster if the mass of planet is lager. Therefore, the difference of the $\delta P$ caused by different planetary masses is quickly eliminated. Moreover, it is worth noting that for stars with an initial rotation period of 16.0$\,$days and a stellar metallicity of -0.5$\,$dex, the spin-up caused by planetary engulfment has a great influence during the whole main sequence, with a period change rate always exceeding 20$\%$. In order to better explain this phenomenon, we introduce panels (a) and (b) in  Figure \ref{fig:mpini}. It can be seen from the  Figure \ref{fig:mpini} that when the planets of 1.0$\,$$M_{J}$, 8.0$\,$$M_{J}$ and 13.0$\,$$M_{J}$ are engulfed, the the angular frequency of stars with different metallicities converges uniformly within 2.0$\,$Gyr. The angular frequency of the stellar rotation has a steep increase when the planet is engulfed, but this increased angular momentum will have an uncontrolled loss. The greater the mass of the planet, the more severe the uncontrolled loss. $\delta P$ will soon be eliminated by the magnetic braking effect. At the same time, comparing the different metallicities, it can be seen that the angular frequency of rotation of metal-poor stars is more difficult to return to the state of stars without planets in the main sequence stage compared to the solar metallicity stars after planets is engulfed. This means that if you consider the situation where stars are polluted by planets, due to the uncertainty of the initial rotation period, the gyrochronology still requires caution for metal-poor stars. For the case of an initial semi-major axis of 0.05$\,$au, only the 13.0$\,$$M_{\mathrm{J}}$ planet is engulfed during the main sequence stage of the star. However, for the case of 0.10$\,$au, due to the sufficient distance which weakens tidal interactions, the orbital decay is slow, and all planetary masses survive. We can clearly observe that when the planet is engulfed and the age is fixed, a larger planetary mass corresponds to a larger $\delta P$.

\subsection{The effects of planetary mass} \label{subsec:planet}

We can also see that for the case of an initial semi-major axis of 0.02$\,$au, as discussed in Section \ref{subsec:axis}, due to the close proximity of the planets to the host star, all planetary masses are engulfed within 1.0$\,$Gyr. For semi-major axis is 0.05$\,$au, With the evolution of time, we find that $\delta P$ has an overall rise. This is due to the transfer of angular momentum caused by the inward migration of the planet to the star, and it can be seen from the magnetic braking formula that under certain conditions, faster stellar rotation will bring about greater angular momentum loss, but the increased angular momentum of the star is always greater than the lost angular momentum, which makes $\delta P$ always increased. And from the simulation results of the planet with a mass of 13.0$\,$$M_{J}$, it's obvious that when the planet is engulfed, $\delta P$ tends to peak, because at this time the planet transfers all the angular momentum to the star. At the same time, the relationship between metallicity and $\delta P$ also shows obvious regularity, that is, the higher the metallicity, the greater the $\delta P$. However, there is little correlation between $\delta P[\% P_{rot,without}]$ and the metallicity of stars. Here we define $\delta P[\% P_{rot,without}] = \delta P/P_{without}$. Except for metal-poor stars with slow initial rotation, different initial rotation periods seem to have little effect on the stellar $\delta P$ and $\delta P[\% P_{rot,without}]$.
For the case where the initial semi-major axis is 0.10$\,$au, it is common for systems with larger planetary masses that the influence of planets on the rotation of stars is greater. For the system with a planetary mass of 1.0$\,$$M_{J}$ until the end of the main sequence,  $\delta P$ is only 0.08$\,$days, which is almost negligible. In the case of more massive planets, $\delta P$ can attain 10.0$\,$days. However, the $\delta P[\% P_{rot,without}]$ of most samples are smaller than 20\%.

\subsection{The effects of stellar mass} \label{subsec:stellar}

\begin{figure}
\centering
\includegraphics[width=\textwidth, angle=0]{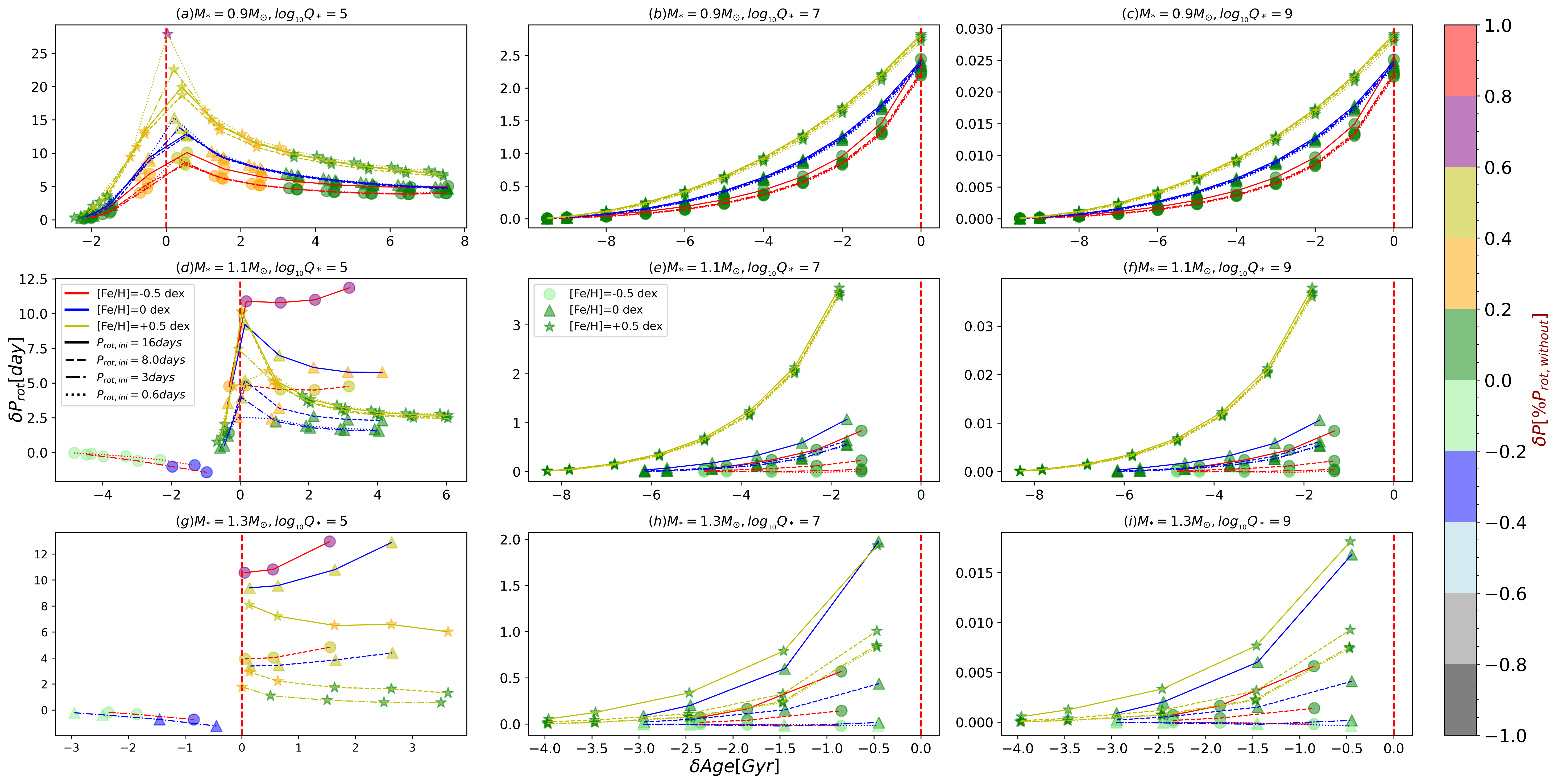}
\caption{The relationship between $\delta Age$ and $\delta P$ of the stellar mass of 0.9, 1.1, 1.3$\,$$M_{\odot}$. Tidal quality parameter $log_{_{10}}\mathcal{Q'}_{_*}$ of 5, 7, 9, initial semi-major axis of 0.05$\,$au, and planetary mass of 1.0$\,$$M_{\mathrm{J}}$. Other parameters are the same as in Figure \ref{fig:1.0}.
\label{fig:0913}}
\end{figure}

\begin{figure*}
  \begin{minipage}[t]{0.5\linewidth}
  \centering
    \caption*{(a)$M_{*}$ = 1.0$\,$$M_{\odot}$,$[Fe/H]$ = -0.5$\,$dex,$a_{\mathrm{ini}}$ = 0.02$\,$au}
   \includegraphics[width=70mm]{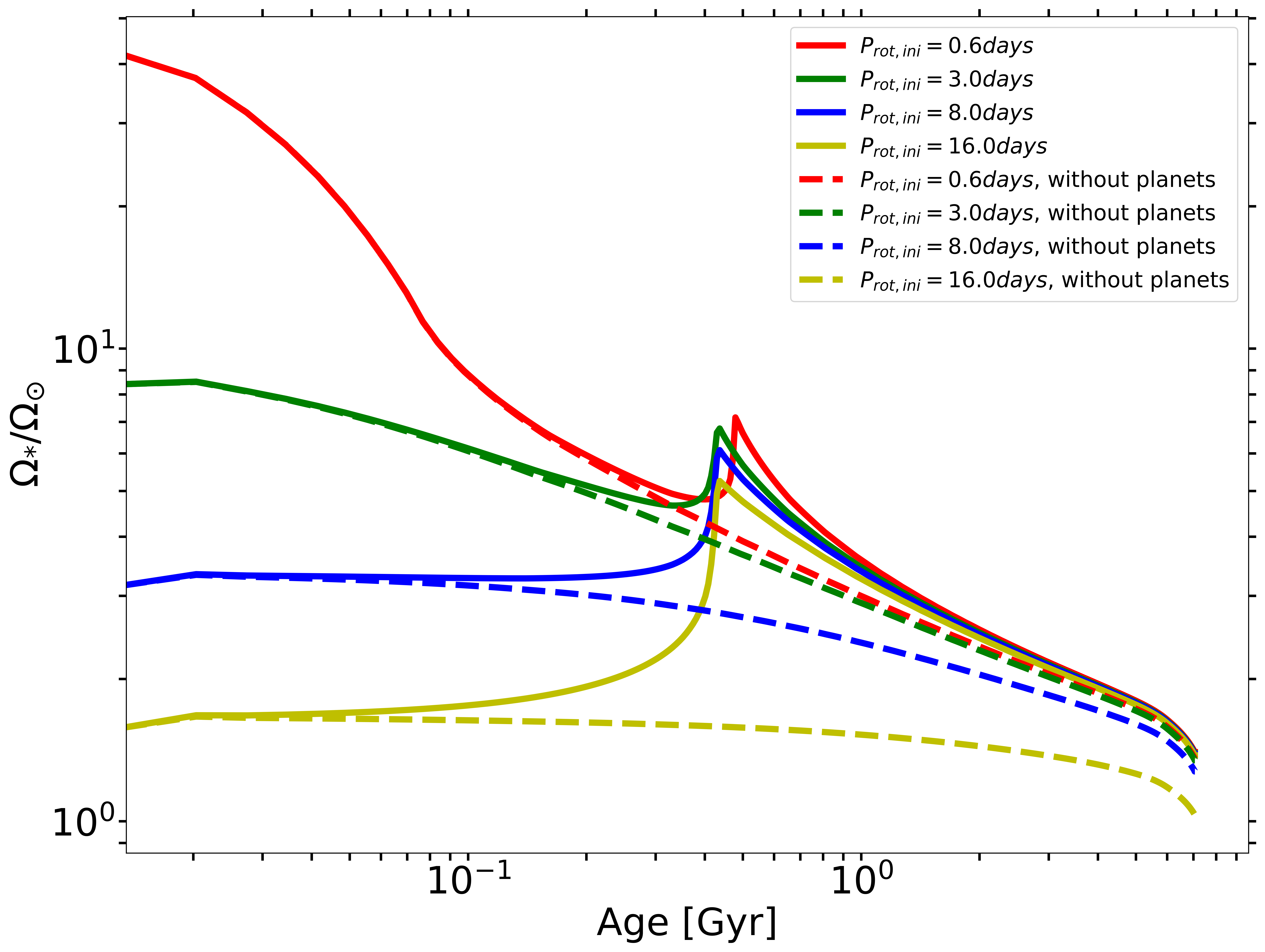}
  \end{minipage}%
  \begin{minipage}[t]{0.5\linewidth}
  \centering
    \caption*{(b)$M_{*}$ = 1.0$\,$$M_{\odot}$,$[Fe/H]$ = +0.5$\,$dex,$a_{\mathrm{ini}}$ = 0.02$\,$au}
   \includegraphics[width=70mm]{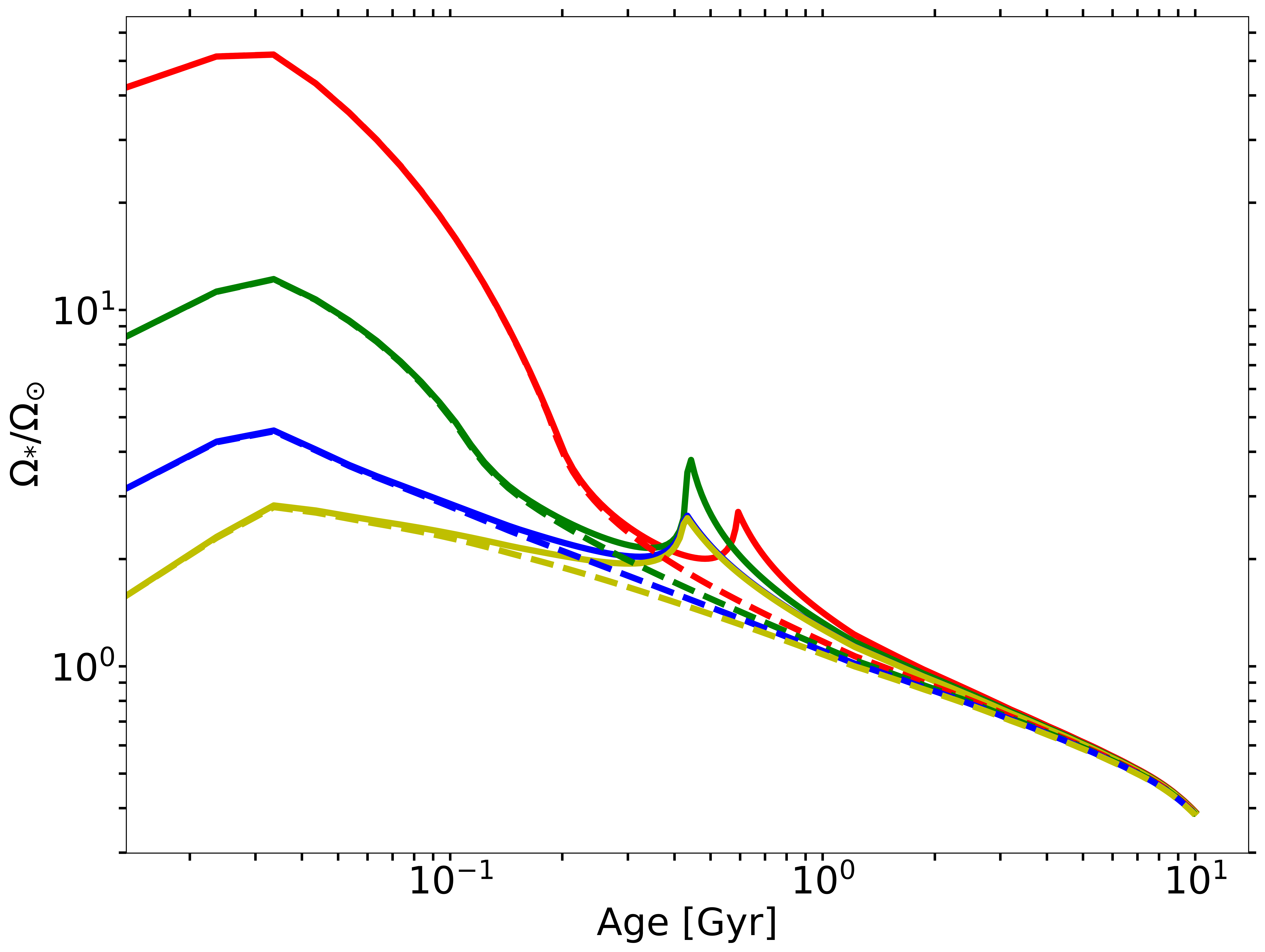}
  \end{minipage}  \\
  \begin{minipage}[t]{0.5\linewidth}
  \centering
    \caption*{(c)$M_{*}$ = 1.3$\,$$M_{\odot}$,$[Fe/H]$ = -0.5$\,$dex,$a_{\mathrm{ini}}$ = 0.02$\,$au}
   \includegraphics[width=70mm]{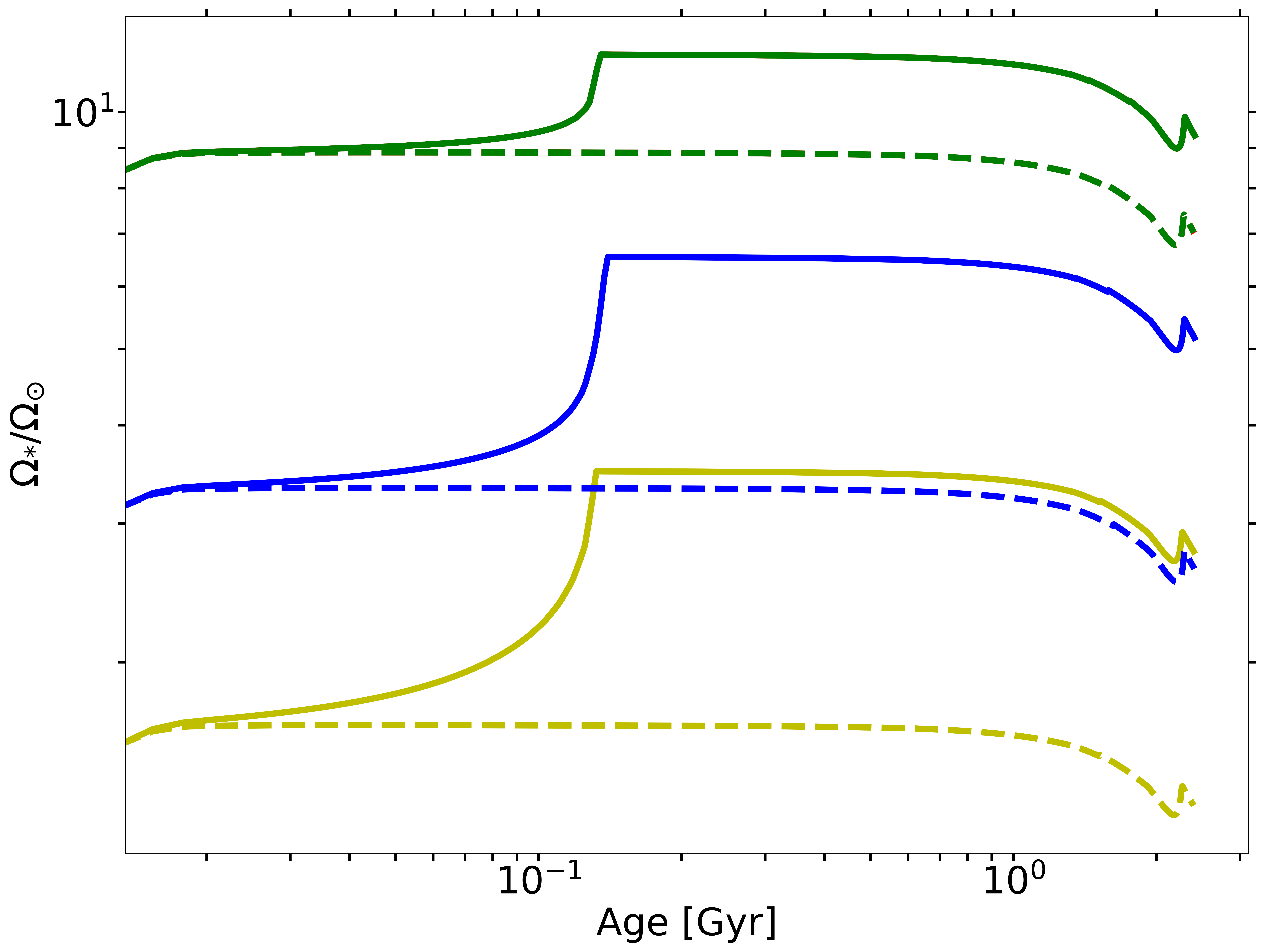}
  \end{minipage}%
  \begin{minipage}[t]{0.5\linewidth}
  \centering
    \caption*{(d)$M_{*}$ = 1.3$\,$$M_{\odot}$,$[Fe/H]$ = +0.5$\,$dex,$a_{\mathrm{ini}}$ = 0.02$\,$au}
   \includegraphics[width=70mm]{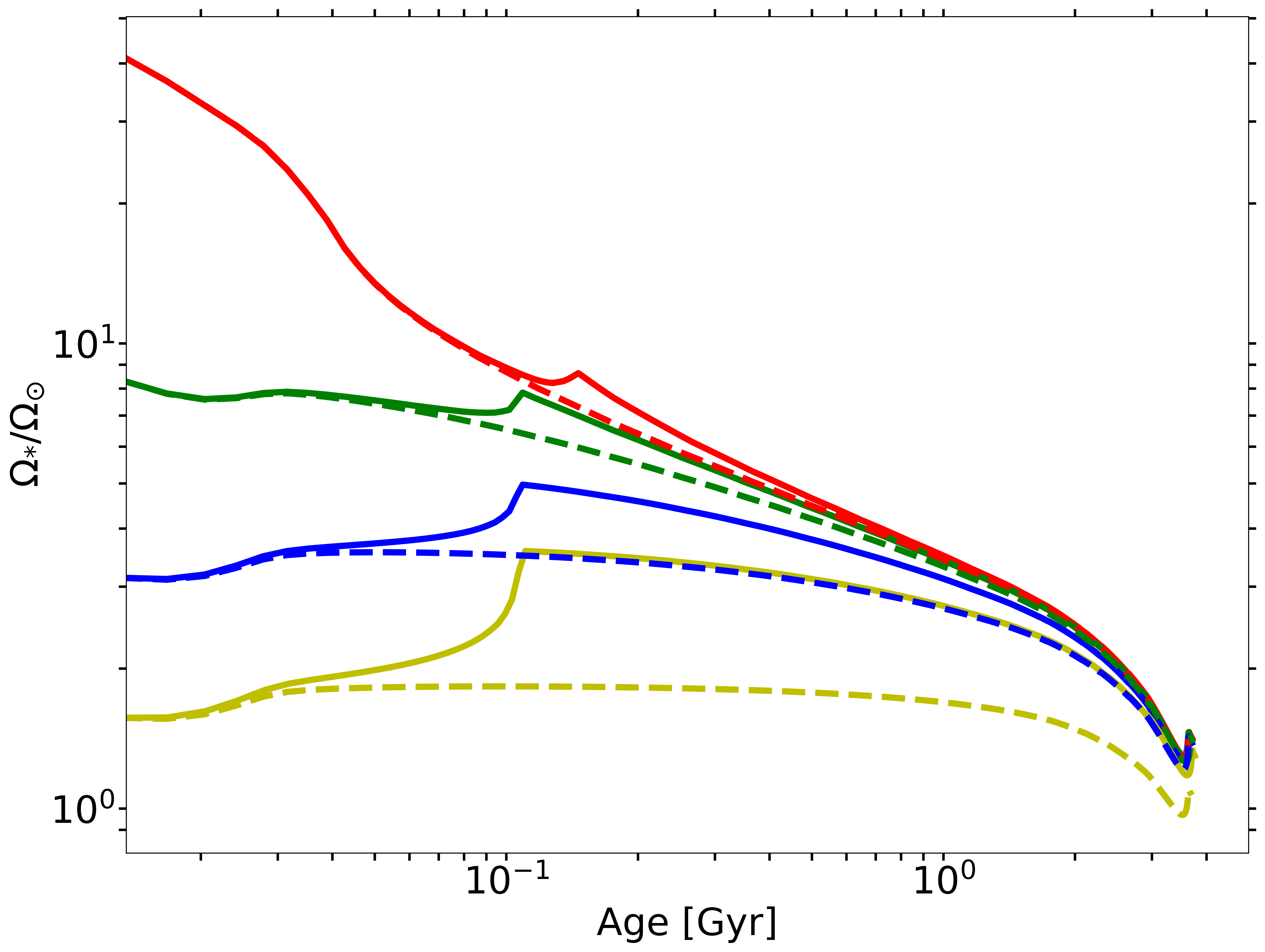}
  \end{minipage}%

\caption{Evolution of the rotation of star as a function of time. The mass of the star is 1.0$\,$$M_{\odot}$ and 1.3$\,$$M_{\odot}$, respectively. The metallicity [Fe/H] is -0.5$\,$dex and +0.5$\,$dex, respectively. The initial semi-major axis is 0.02 au, and the mass of the planet is 1.0$\,$$M_{\mathrm{J}}$. The tidal quality parameter $log_{_{10}}\mathcal{Q'}_{_*}$$\,$ = $\,$7. The dashed line in the figure is the model without planets, and the solid line is the model with planets. Red, green, blue, and yellow indicate the initial rotation period of the star is 0.6, 3.0, 8.0, and 16.0$\,$days, respectively.
\label{fig:pini}}
\end{figure*}

\begin{figure*}
  \begin{minipage}[t]{0.45\linewidth}
  \centering
    \caption*{(a)$M_{*}$ = 1.0$\,$$M_{\odot}$,$[Fe/H]$ = -0.5$\,$dex,$a_{\mathrm{ini}}$ = 0.02$\,$au}
   \includegraphics[width=50mm]{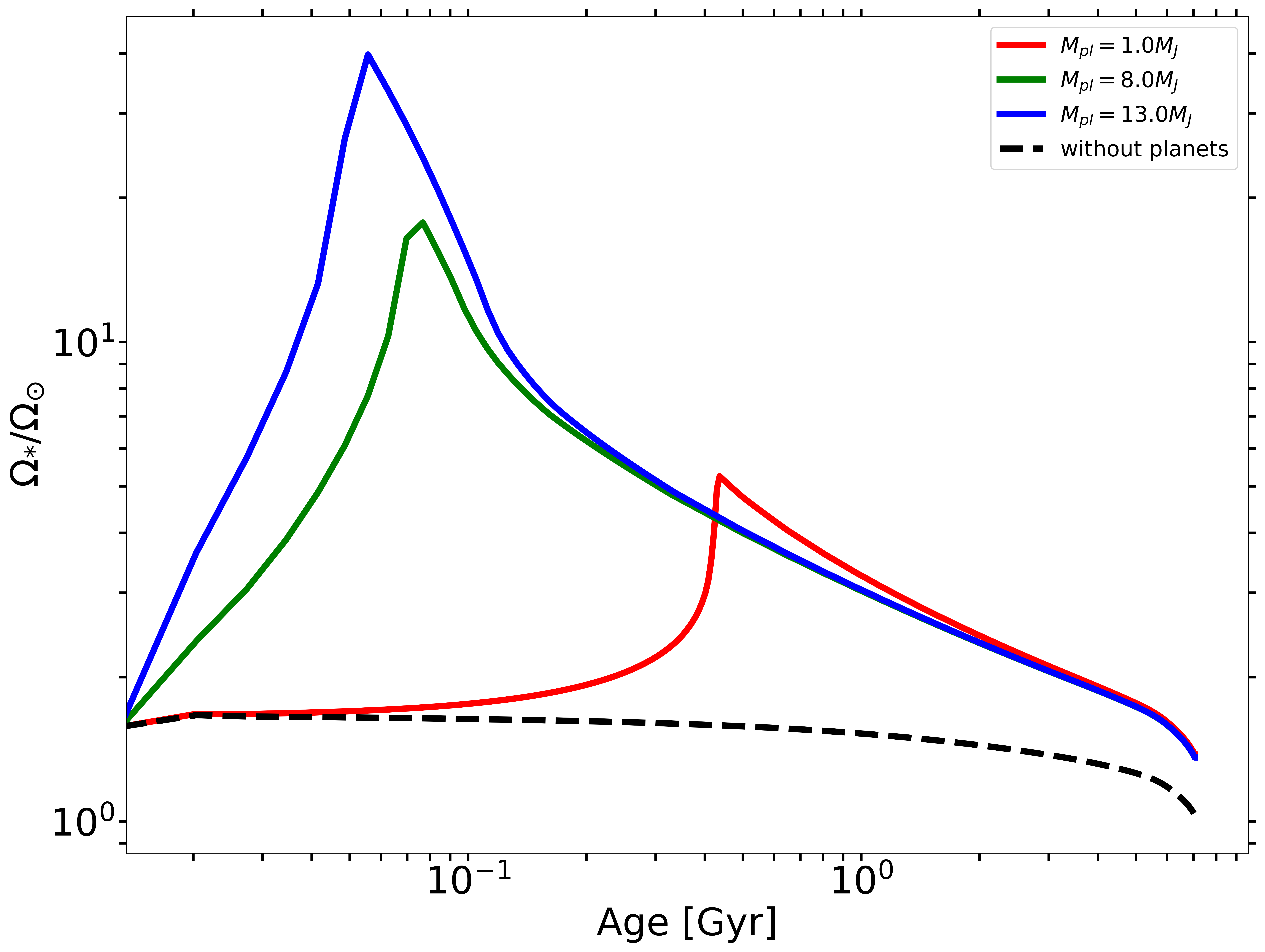}
  \end{minipage}%
  \begin{minipage}[t]{0.45\linewidth}
  \centering
    \caption*{(b)$M_{*}$ = 1.0$\,$$M_{\odot}$,$[Fe/H]$ = +0.5$\,$dex,$a_{\mathrm{ini}}$ = 0.02$\,$au}
   \includegraphics[width=50mm]{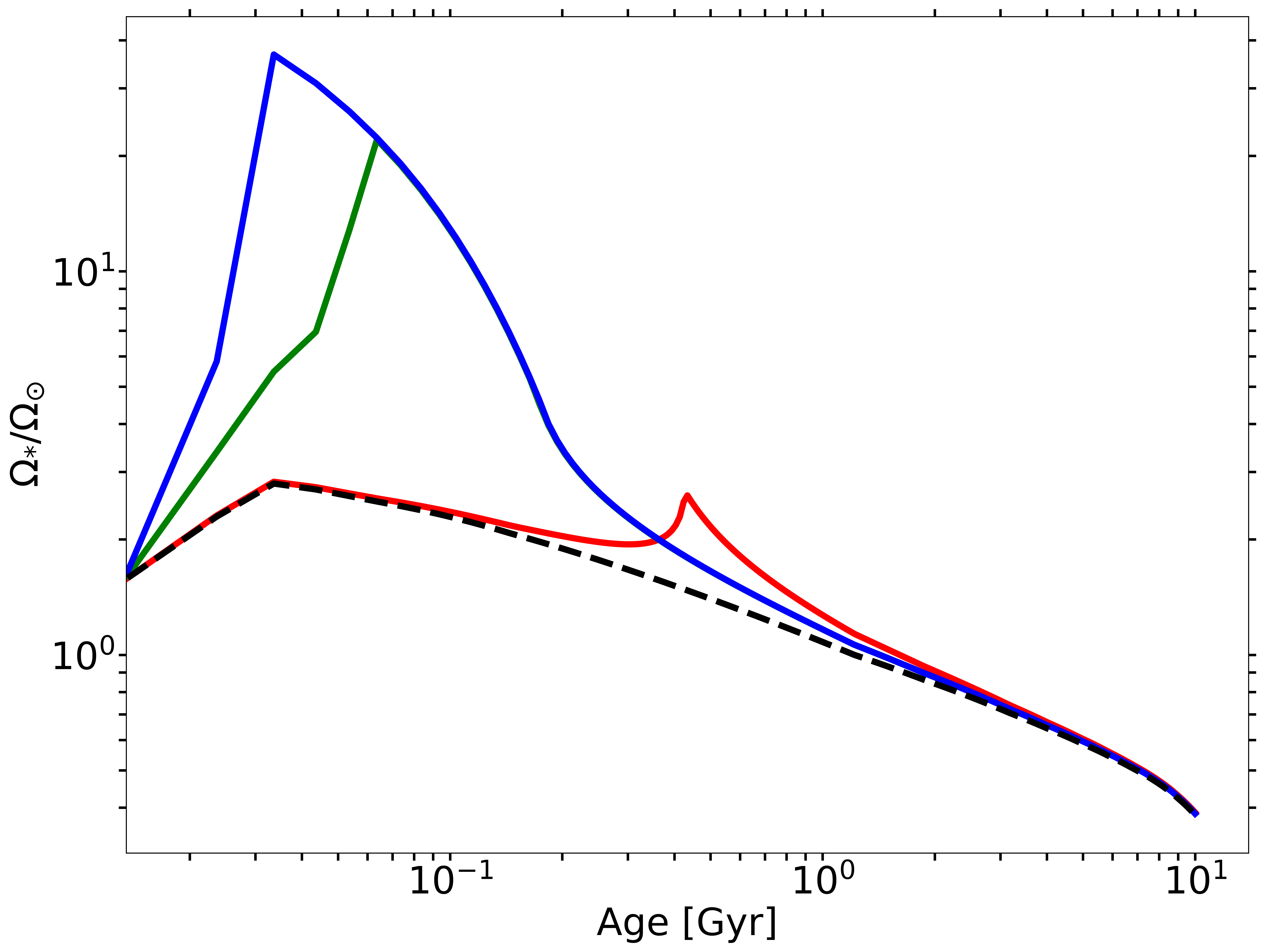}
  \end{minipage}  \\
  \begin{minipage}[t]{0.45\linewidth}
  \centering
    \caption*{(c)$M_{*}$ = 1.0$\,$$M_{\odot}$,$[Fe/H]$ = -0.5$\,$dex,$a_{\mathrm{ini}}$ = 0.03$\,$au}
   \includegraphics[width=50mm]{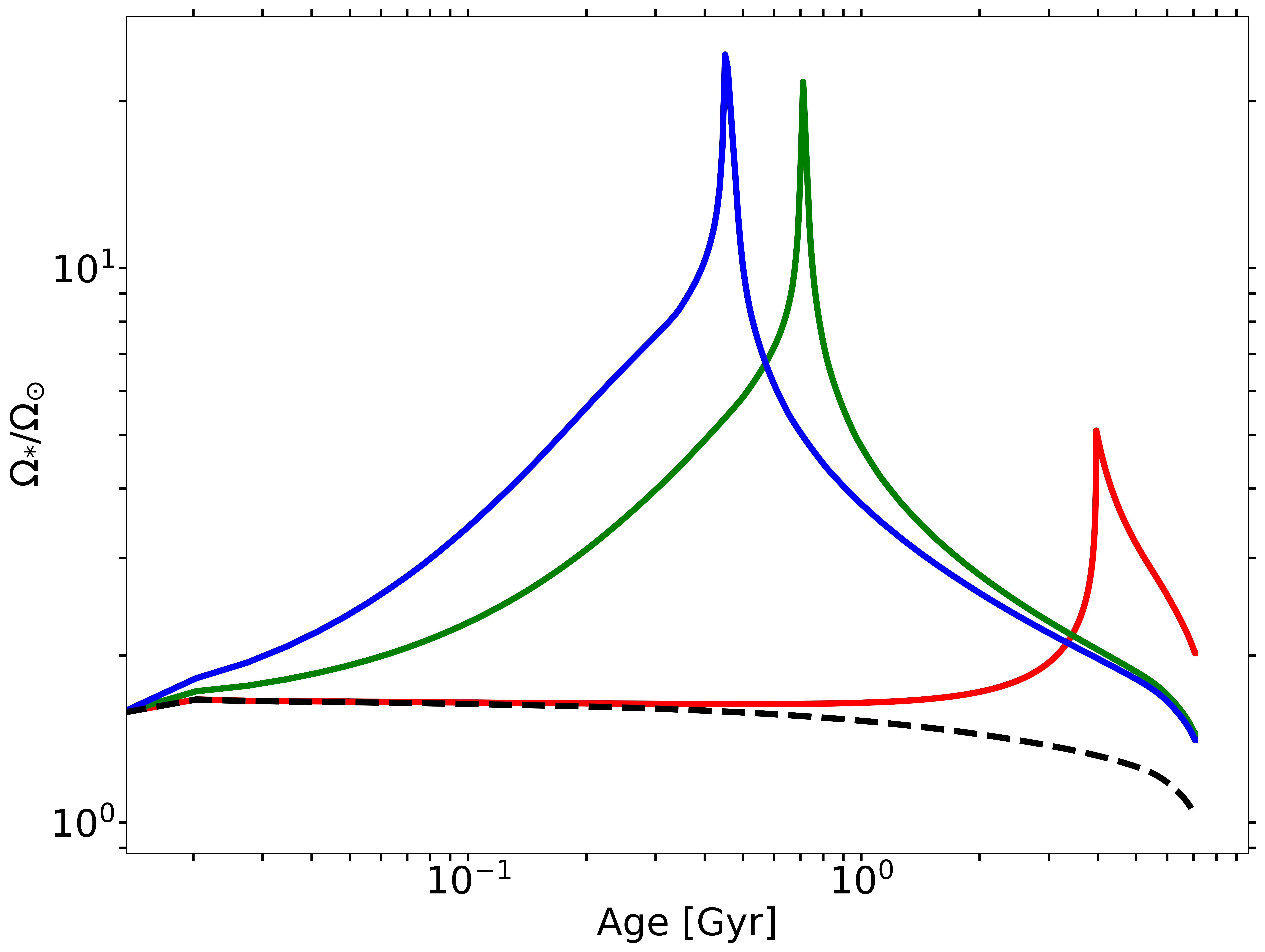}
  \end{minipage}%
  \begin{minipage}[t]{0.45\linewidth}
  \centering
    \caption*{(d)$M_{*}$ = 1.0$\,$$M_{\odot}$,$[Fe/H]$ = +0.5$\,$dex,$a_{\mathrm{ini}}$ = 0.03$\,$au}
   \includegraphics[width=50mm]{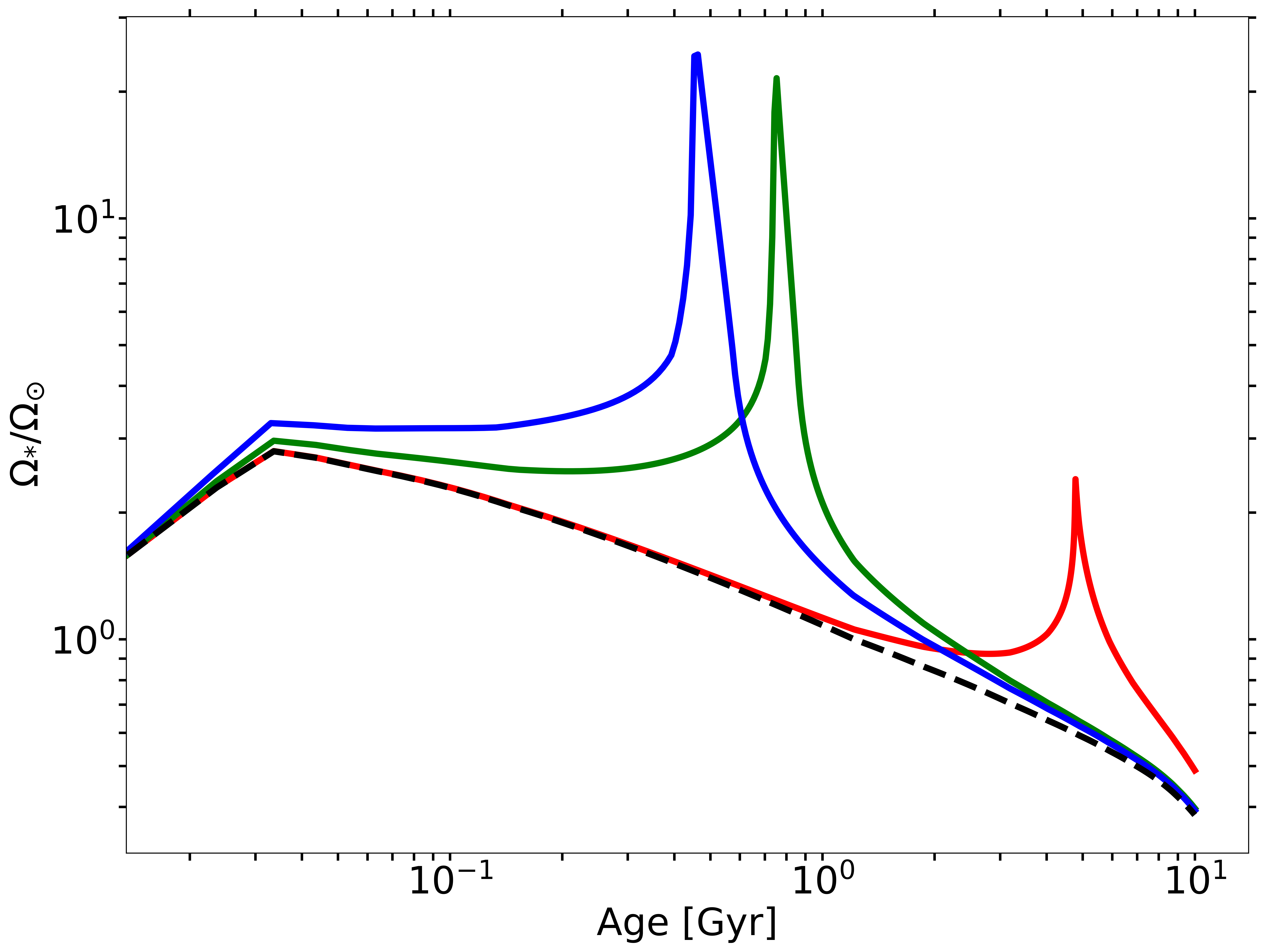}
  \end{minipage}  \\
  \begin{minipage}[t]{0.45\linewidth}
  \centering
    \caption*{(e)$M_{*}$ = 1.3$\,$$M_{\odot}$,$[Fe/H]$ = -0.5$\,$dex,$a_{\mathrm{ini}}$ = 0.02$\,$au}
   \includegraphics[width=50mm]{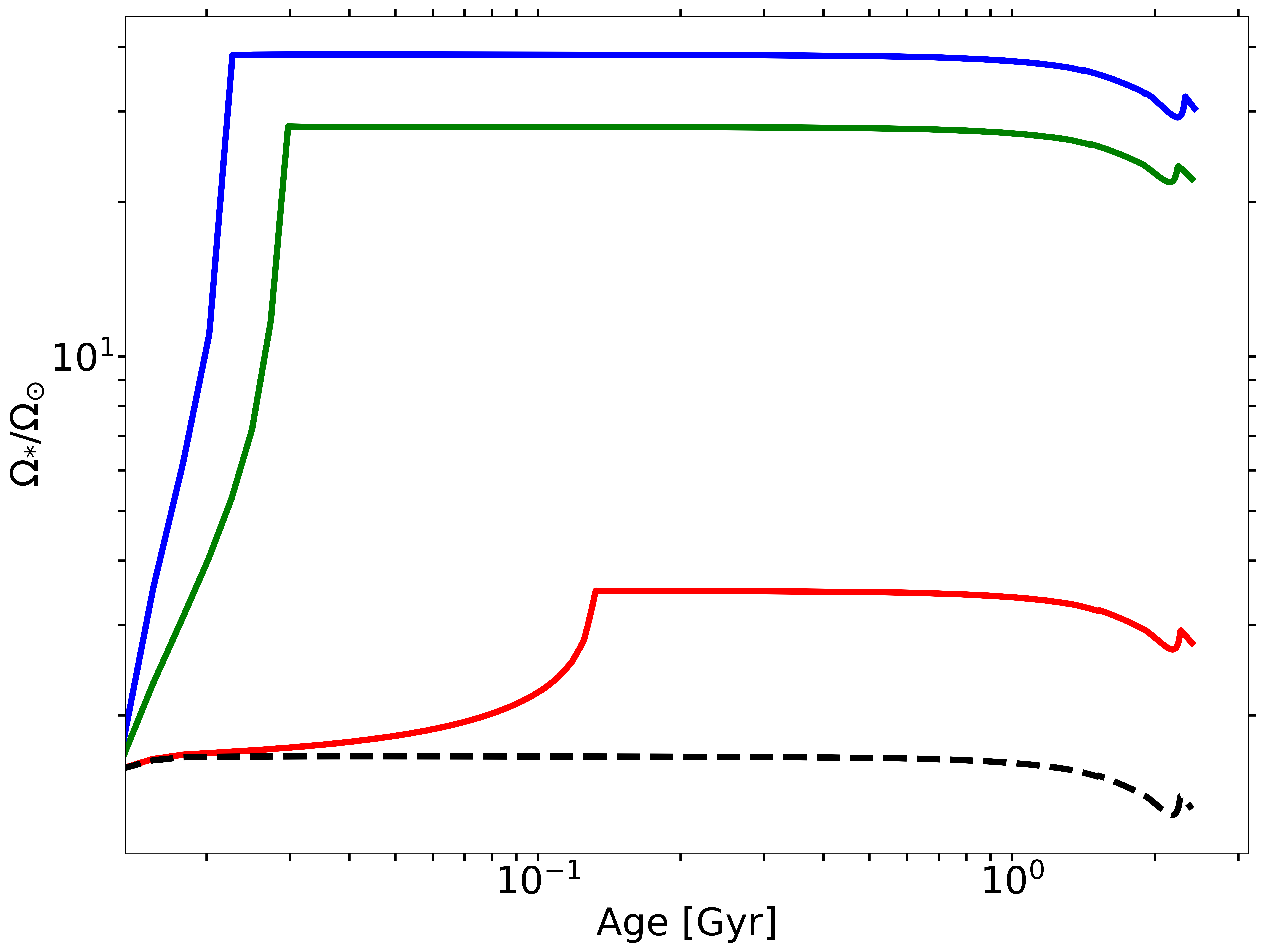}
  \end{minipage}%
  \begin{minipage}[t]{0.45\linewidth}
  \centering
    \caption*{(f)$M_{*}$ = 1.3$\,$$M_{\odot}$,$[Fe/H]$ = +0.5$\,$dex,$a_{\mathrm{ini}}$ = 0.02$\,$au}
   \includegraphics[width=50mm]{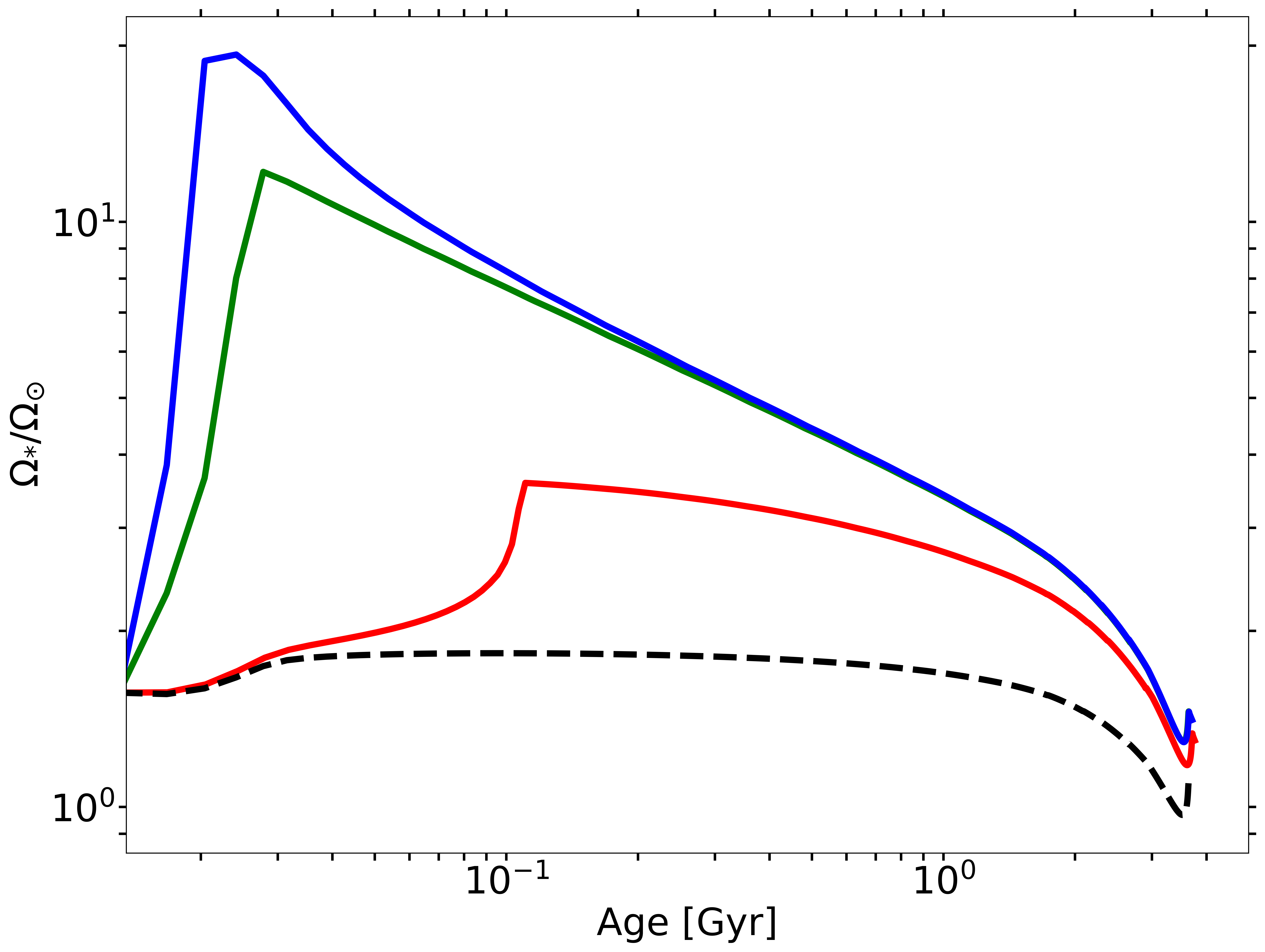}
  \end{minipage}  \\
  \begin{minipage}[t]{0.45\linewidth}
  \centering
    \caption*{(g)$M_{*}$ = 1.3$\,$$M_{\odot}$,$[Fe/H]$ = -0.5$\,$dex,$a_{\mathrm{ini}}$ = 0.03$\,$au}
   \includegraphics[width=50mm]{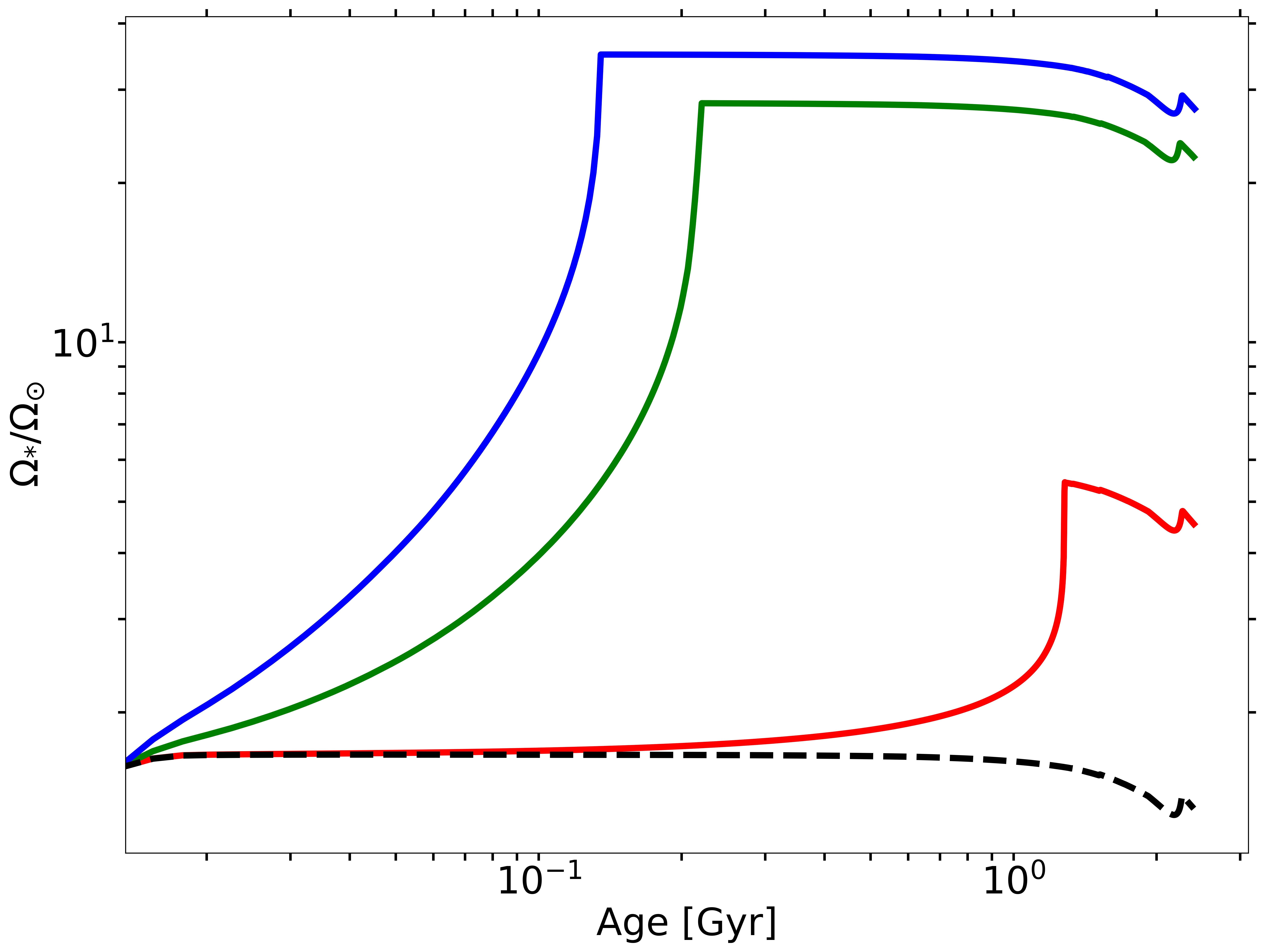}
  \end{minipage}%
  \begin{minipage}[t]{0.45\linewidth}
  \centering
    \caption*{(h)$M_{*}$ = 1.3$\,$$M_{\odot}$,$[Fe/H]$ = +0.5$\,$dex,$a_{\mathrm{ini}}$ = 0.03$\,$au}
   \includegraphics[width=50mm]{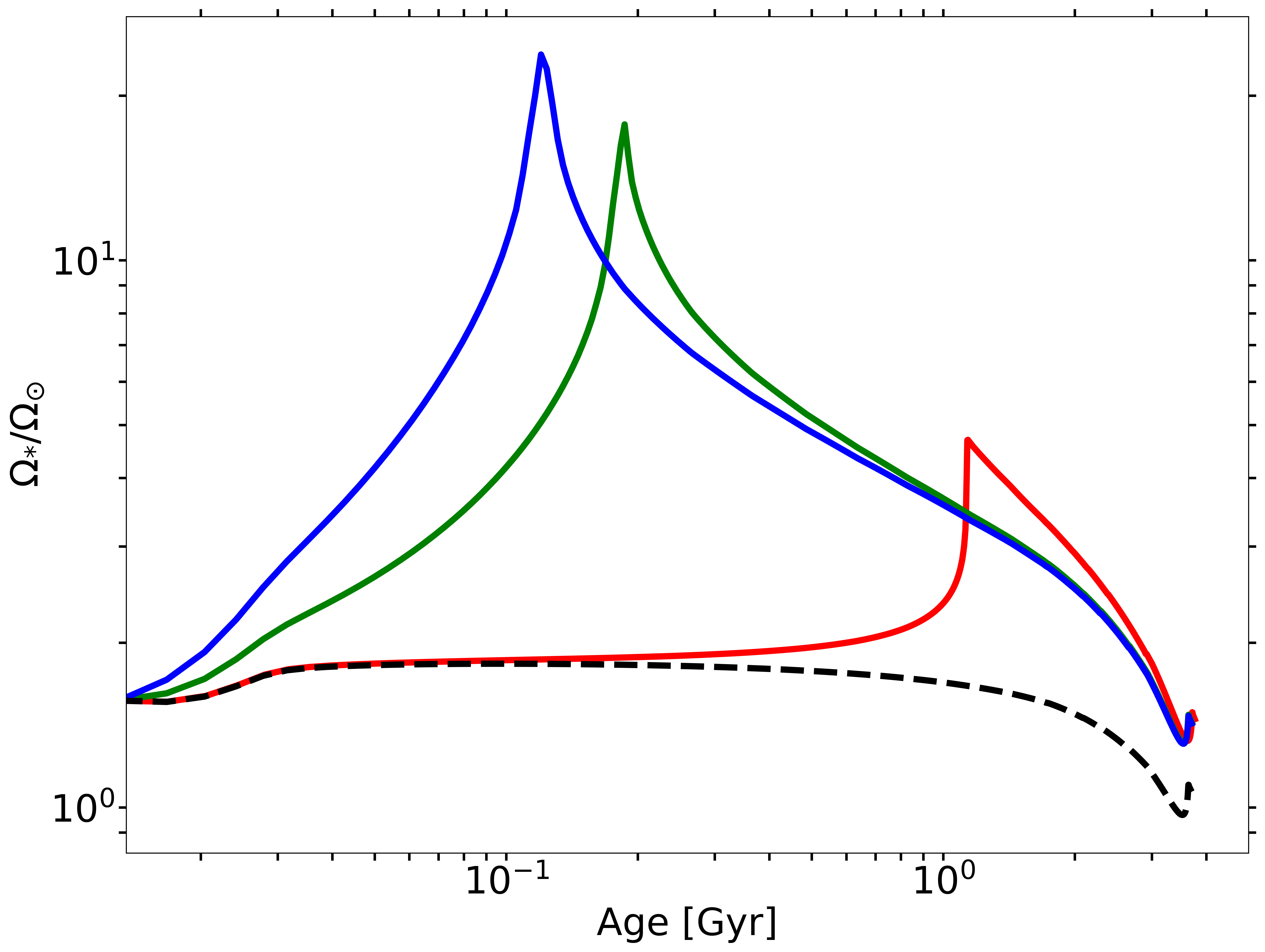}
  \end{minipage}%
\caption{Evolution of the rotation of star as a function of time. The mass of the star is 1.0$\,$$M_{\odot}$ and 1.3$\,$$M_{\odot}$, the metal abundance [Fe/H] is -0.5$\,$dex and +0.5$\,$dex, respectively. The initial semi-major axis is 0.02 and 0.03$\,$au, respectively. The tidal quality parameter $log_{_{10}}\mathcal{Q'}_{_*}$ = 7. And the initial rotation period of the star is 16.0$\,$days. The dotted lines are the models without planets and the solid lines are the models with planets. Red, green, and blue indicate the planets with masses of 1.0, 8.0, and 13.0$\,$$M_{\mathrm{J}}$, respectively.
\label{fig:mpini}}
\end{figure*}

\begin{figure*}
  \begin{minipage}[t]{0.5\linewidth}
  \centering
    \caption*{(a)$M_{*}$ = 1.0$\,$$M_{\odot}$,$[Fe/H]$ = -0.5$\,$dex,$a_{\mathrm{ini}}$ = 0.02$\,$au}
   \includegraphics[width=70mm]{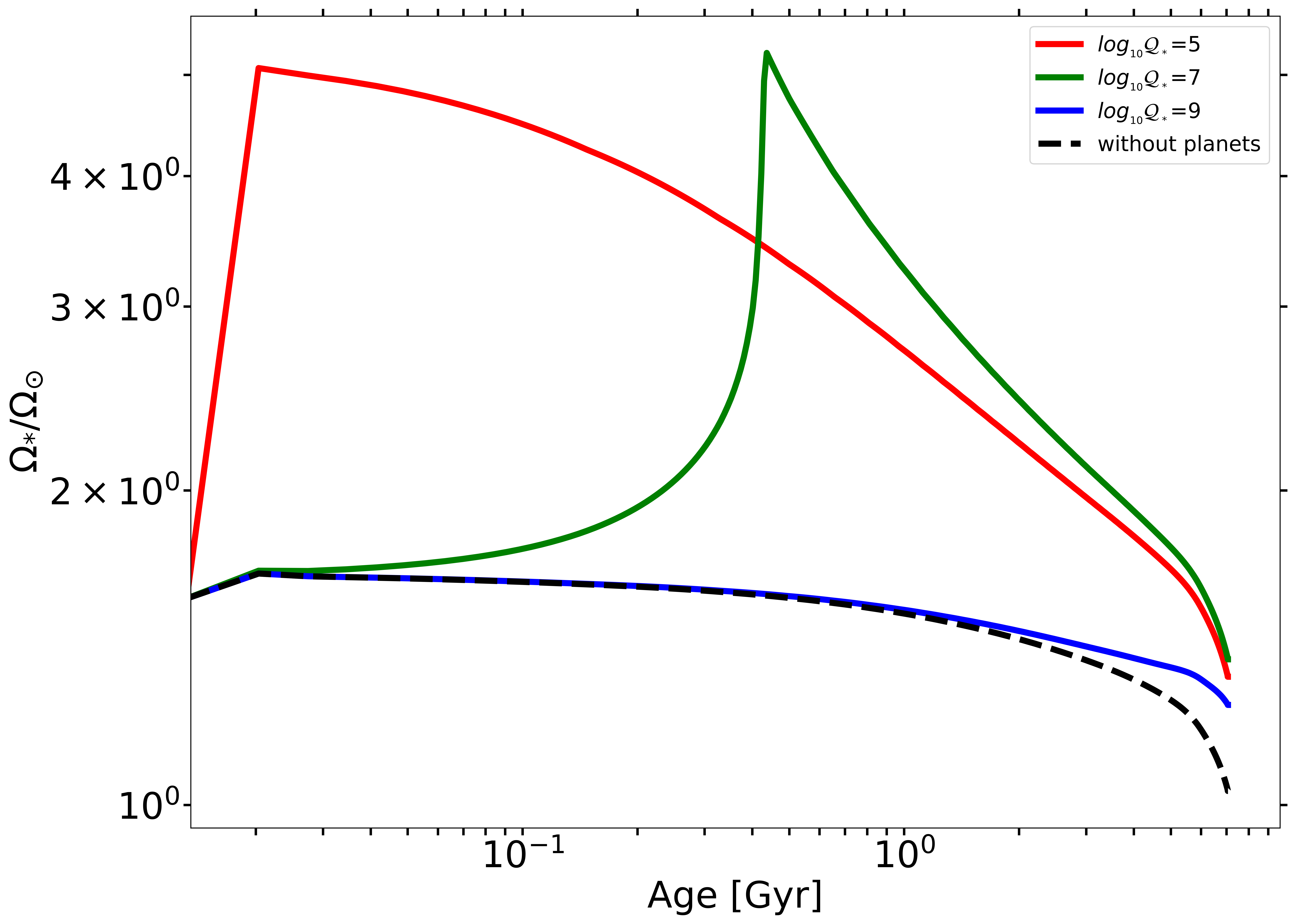}
  \end{minipage}%
  \begin{minipage}[t]{0.5\linewidth}
  \centering
    \caption*{(b)$M_{*}$ = 1.0$\,$$M_{\odot}$,$[Fe/H]$ = +0.5$\,$dex,$a_{\mathrm{ini}}$ = 0.02$\,$au}
   \includegraphics[width=70mm]{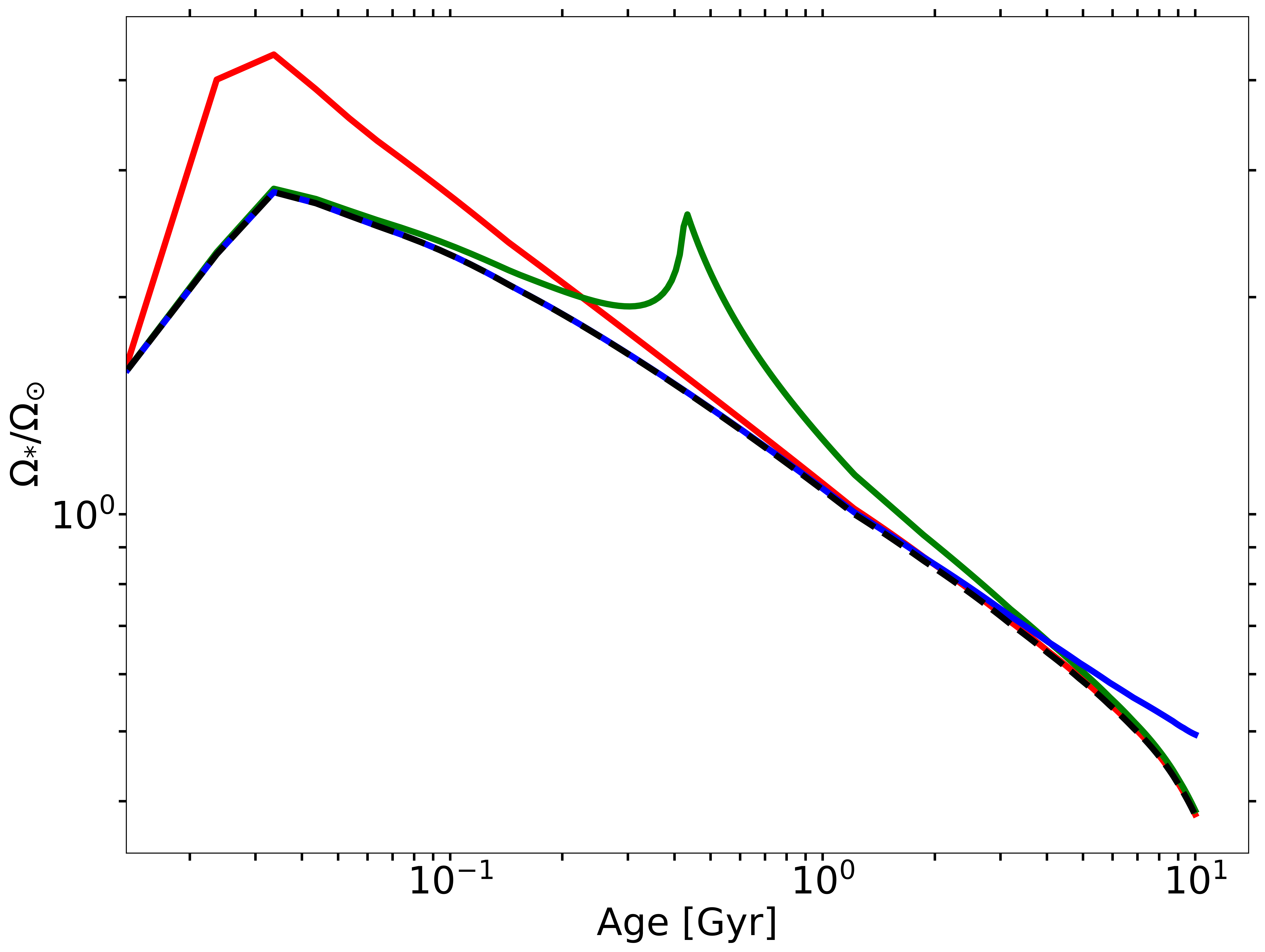}
  \end{minipage}  \\
  \begin{minipage}[t]{0.5\linewidth}
  \centering
    \caption*{(c)$M_{*}$ = 1.3$\,$$M_{\odot}$,$[Fe/H]$ = -0.5$\,$dex,$a_{\mathrm{ini}}$ = 0.02$\,$au}
   \includegraphics[width=70mm]{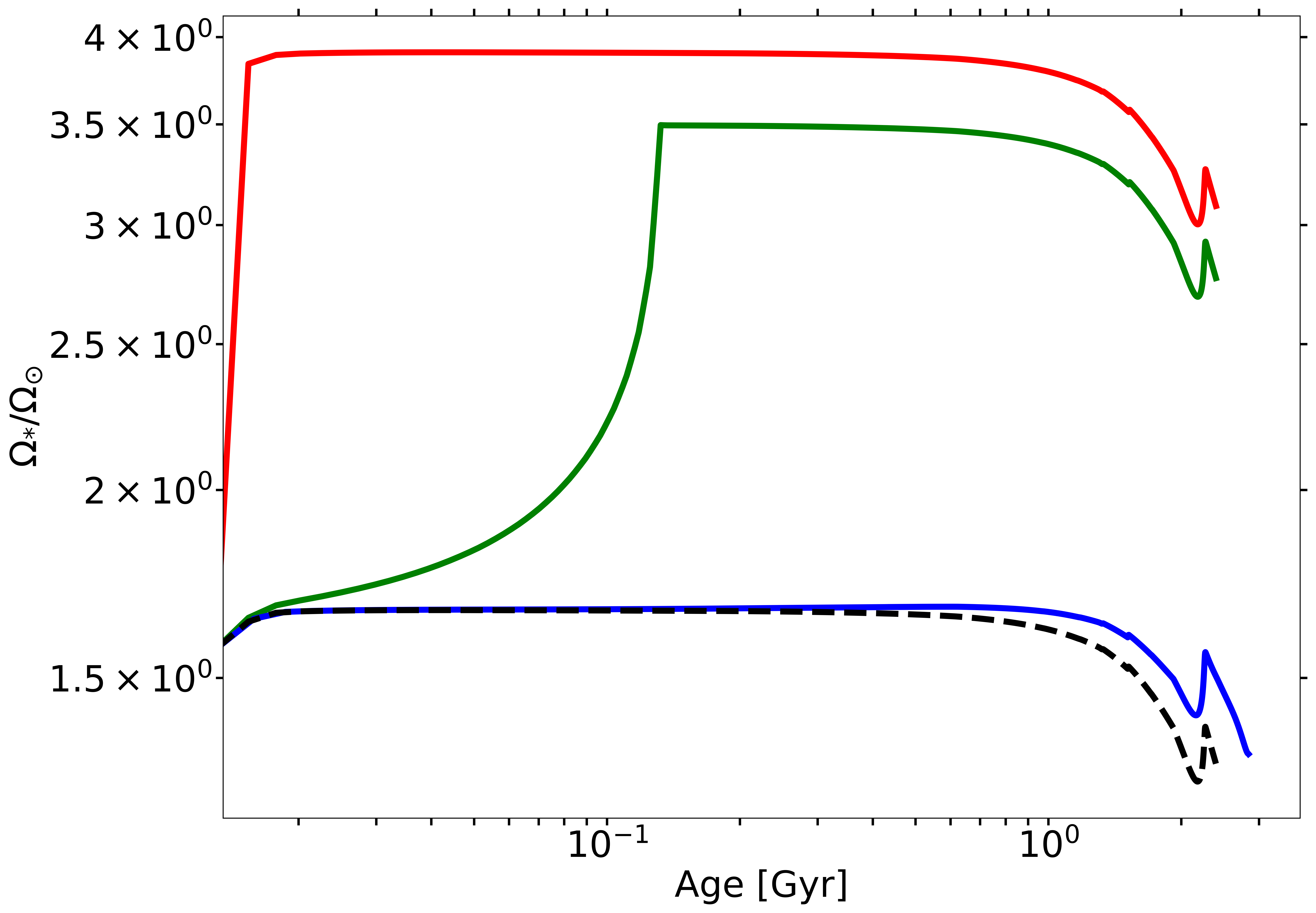}
  \end{minipage}%
  \begin{minipage}[t]{0.5\linewidth}
  \centering
    \caption*{(d)$M_{*}$ = 1.3$\,$$M_{\odot}$,$[Fe/H]$ = +0.5$\,$dex,$a_{\mathrm{ini}}$ = 0.02$\,$au}
   \includegraphics[width=70mm]{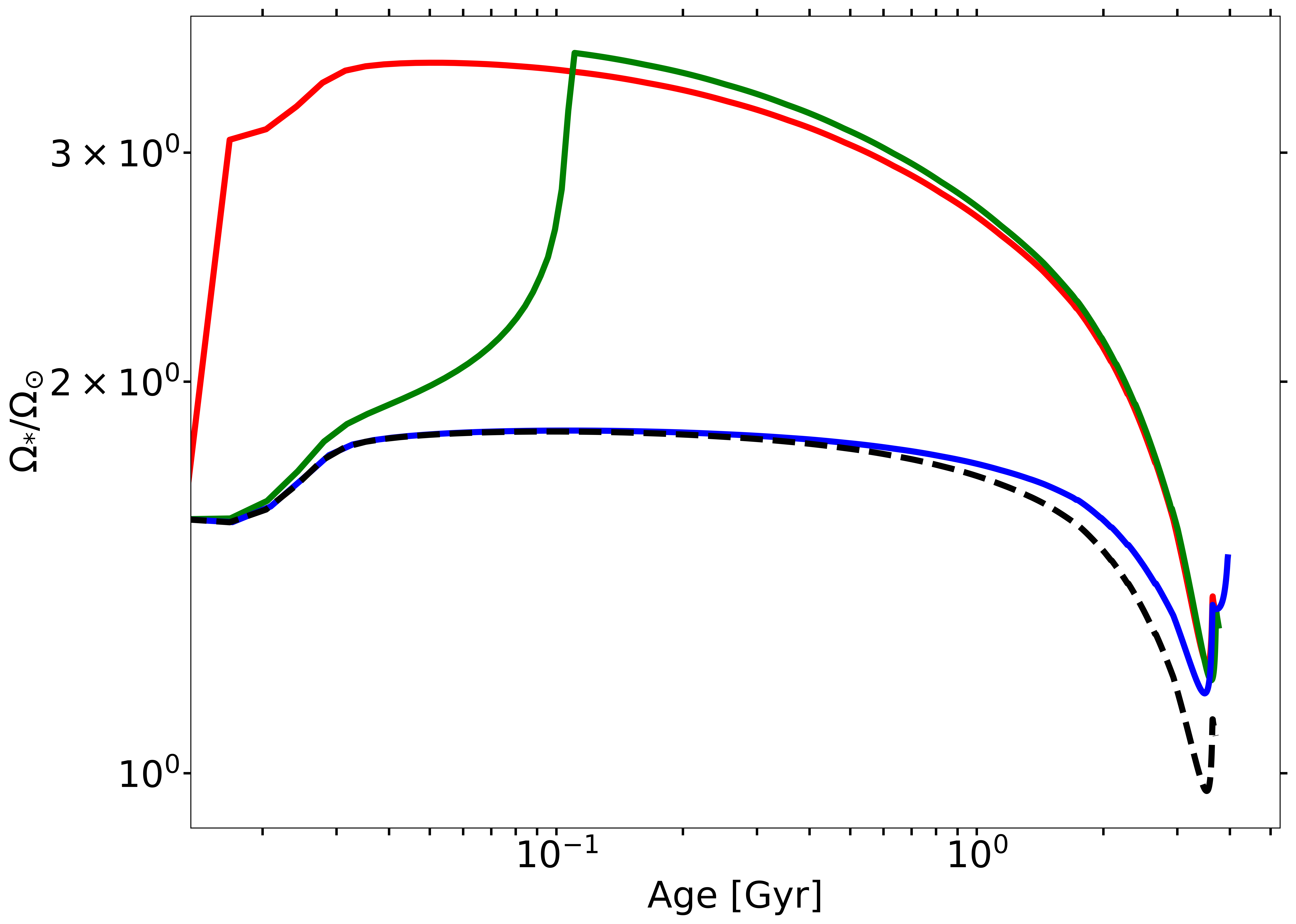}
  \end{minipage}%
\caption{Evolution of the rotation of star as a function of time. The mass of the star is 1.0$\,$$M_{\odot}$ and 1.3$\,$$M_{\odot}$, respectively. The metallicity [Fe/H] is -0.5$\,$dex and +0.5$\,$dex, respectively. The initial semi-major axis is 0.02$\,$au, and the mass of the planet is 1.0$\,$$M_{\mathrm{J}}$. The initial rotation period of the star is 16.0$\,$days. The dashed line in the figure is the model without planets, and the solid line is the model with planets. Red, green and blue indicate the tidal quality parameter $log_{_{10}}\mathcal{Q'}_{_*}$ is 5, 7 and 9, respectively.
\label{fig:Q}}
\end{figure*}

We show the variations of stellar periods for the case of 1.0$\,$M$_{\mathrm{J}}$ and initial orbital distance of 0.05$\,$au in Figure \ref{fig:0913}. The engulfment of planet only occurs at $log_{_{10}}\mathcal{Q'}_{_*}$ = 5. The panel (a) of Figure \ref{fig:0913} shows that for the 0.9$\,$$M_{\odot}$, the higher the metallicity, the greater the $\delta P$. This is because the planets transfer angular momentum to compensate the rapid loss of angular momentum for the stars with low metallicities (thin convective envelopes). Thus, $\delta P$ is small. However, for the stars with high metallicities (thick convective envelopes), the angular momentum provided by the planets is insufficient. In this situation, $\delta P$ is large. The effect of metallicity on the stellar periods shows the opposite trend in high mass stars. As shown in panels (d) and (g), $\delta P$ is large in the cases of low metallicity. In fact, the convective envelopes of 1.1 and 1.3$\,$$M_{\odot}$ stars are very thin in the case of low metallicity. The angular momentum provided by the planet can be remaind after the planet is engulfed. For the cases of high metallicity, those angular momentum will be rapidly lost owing to the magnetic braking. Thus, $\delta P$ is smaller than that of low metallicity. The influence of planets on the rotation period of stars is slight ($P_{\mathrm{rot, ini}}$ $<$ 4.0$\,$days) or even negligible during the entire stellar main sequence for the cases of $log_{_{10}}\mathcal{Q'}_{_*}$ = 7 and $log_{_{10}}\mathcal{Q'}_{_*}$ = 9 because of the weak interaction (The middle and right panels of Figure \ref{fig:0913}).

We compare star-planet systems with an initial stellar mass of 1.1$\,$$M_{\odot}$ and 1.3$\,$$M_{\odot}$. It was found that the size of $\delta P$ is different from that of low-mass stars, but it is closely related to the initial stellar rotation period. When the metallicity is smaller and the initial stellar rotation period is larger, stars tend to have larger $\delta P$. And in Figures \ref{fig:0913} (d) and (g), there are two phenomena for 1.1$\,$$M_{\odot}$ metal-poor stars and 1.3$\,$$M_{\odot}$ metal-poor and solar metal-rich stars:
One is that for systems with an initial rotation period of less than 3.0$\,$days, the planets have survived and $\delta P$ is reduced and negative, and the angular momentum is transferred from the star to the planetary orbit. At this time, $\Omega$ $\geq$ n and the planet is always in tidal migration outward. Another phenomenon is that after the planet is engulfed, $\delta P$ increases for the initial stellar rotation period of 16.0 days and 8.0$\,$days. This is opposite for the  stars with smaller masses or higher metallicities, because after the planet is engulfed, the star only loses angular momentum by magnetic braking, and there is no angular momentum transfer between the planet's orbit and the star. From  Figures \ref{fig:pini} (c) and (d), it can be seen that stars with a mass of 1.3$\,$$M_{\odot}$ and a metallicity of -0.5$\,$dex will not experience significant angular momentum loss until the end of the main sequence under any initial rotation period, and the increase in stellar rotation caused by the planet being engulfed will also continue almost lossless until the end of the main sequence. In Figures \ref{fig:pini} (c), we do not have the portion corresponding to $P_{rot}$ = 0.6$\,$days. This is because for a star with a metallicity of [Fe/H] = -0.5$\,$dex and a mass of 1.3$\,$$M_{\odot}$, the rapid rotation model with a period of 0.6$\,$days exceeds the critical rotational velocity of the star due to its initial larger radius. Therefore, MESA is unable to generate such models. In the case of stellar metallicity of +0.5$\,$dex, except for the case where the stellar initial rotation is too slow. When the planet is engulfed, during the main sequence, for the stars with different initial periods, the influence of planetary engulfment on the rotation of stars will still be eliminated.
We can summarize the following rules from the 1.3$\,$$M_{\odot}$ star in  Figure \ref{fig:mpini}:
(\romannumeral1) Metal-rich stars have thicker convective envelopes, and for longer orbital migrations, slower planetary engulfment will also bring about slower $\delta P$ elimination.
(\romannumeral2) Since the convective envelopes of metal-poor stars are almost negligible, magnetic braking will hardly cause angular momentum loss, and the speed of planetary engulfment does not affect $\delta P$.

\subsection{The effects of tidal quality parameter} \label{subsec:quality}

In Section \ref{subsec:stellar}, we extensively discuss the evolution of $\delta P$ for different stellar masses under the condition of $log_{10}\mathcal{Q'}_{_*}$ = 5. Throughout the main sequence stage, planets survive under the cases of $log{10}\mathcal{Q'}_{_*}$ = 7 and $log{10}\mathcal{Q'}_{_*}$ = 9, with the influence of stellar-planet interactions on the stellar rotation period being weak ($\delta P$ $<$ 4.0$\,$days) or even negligible (as shown in the middle and right panels of Figure \ref{fig:0913}). Furthermore, although these two tidal dissipation factors have different values for $\delta P$, their trends are almost identical. Finally, we can see from  Figure \ref{fig:Q} that the tidal quality parameter $Q'_{*}$ will have a great impact on the timescale of planetary migration, and with $log_{_{10}}\mathcal{Q'}_{_*}$ = 9, Jupiter-mass planets have survived during the main sequence. Comparing the two different metallicity situations when the mass of the star is 1.0$\,$$M_{\odot}$, we can see that metal-rich stars will still experience faster stellar rotation angle frequency loss after the planet is swallowed and stellar rotation angle frequency converge consistently during the main sequence. For stars with a mass of 1.3$\,$$M_{\odot}$ and metallicity of -0.5$\,$dex, the spin-up of the star caused by the planet being swallowed will also continue almost lossless until the end of the main sequence.

\subsection{The evolution of WASP-19} \label{Wasp-19}

In this section, we will use an interaction model to simulate the possible evolutionary trajectory of WASP-19, the parameters of the WASP-19 we used are shown in Table \ref{tab:tab2}. We use the observed age, current stellar rotation period, and current orbital semimajor axis to constrain the initial semimajor axis and tidal quality parameter $Q'_{*}$ of the system. We selected this star-planet system for three main reasons: 
(\romannumeral1) The stellar mass of this system is small and has a thick convective envelope. In our previous discussions, we know that for low-mass metal-rich stars, magnetic braking produces stronger angular momentum loss in the case of a thicker convective envelope. When the evolution timescale is long, the difference in initial stellar rotation periods will be quickly smoothed out over time. The uncertainty in the initial rotation period will not have a significant impact on the evolution of this system. 
(\romannumeral2) The observation error of the stellar rotation period in this system is small, and the interaction model can better constrain the parameters. 
(\romannumeral3) Recent studies on Wasp-19 have measured the orbital decay of the system and calculated the $Q'_{*}$ value \citep{2020AJ....159..150P,2020MNRAS.491.1243P,2022A&A...668A.114R}, this is significant for our results and provides valuable reference.

\begin{table}
\begin{center}
\caption[]{ Parameters of the WASP-19 planetary system \citep{2020A&A...636A..98C}}\label{tab:tab2}
 \begin{tabular}{cl}
  \hline\noalign{\smallskip}
planet WASP-19 &  star WASP-19                  \\
  \hline\noalign{\smallskip}
Mass:1.154$\,$$M_{\mathrm{J}}$  & Mass:0.965$\,$$M_{\odot}$ \\ 
Semi major axis:0.01652 au  & Age: 2.9-10.5$\,$billion years \\ 
--  & Metallicity([Fe/H]):0.04$\,$dex\\ 
--  &  Rotation period:10.3-10.7$\,$days\\ 
  \noalign{\smallskip}\hline
\end{tabular}
\end{center}
\end{table}

\begin{figure}
  \begin{minipage}[t]{\linewidth}
  \centering
    \caption*{(a)}
   \includegraphics[width=0.8\linewidth]{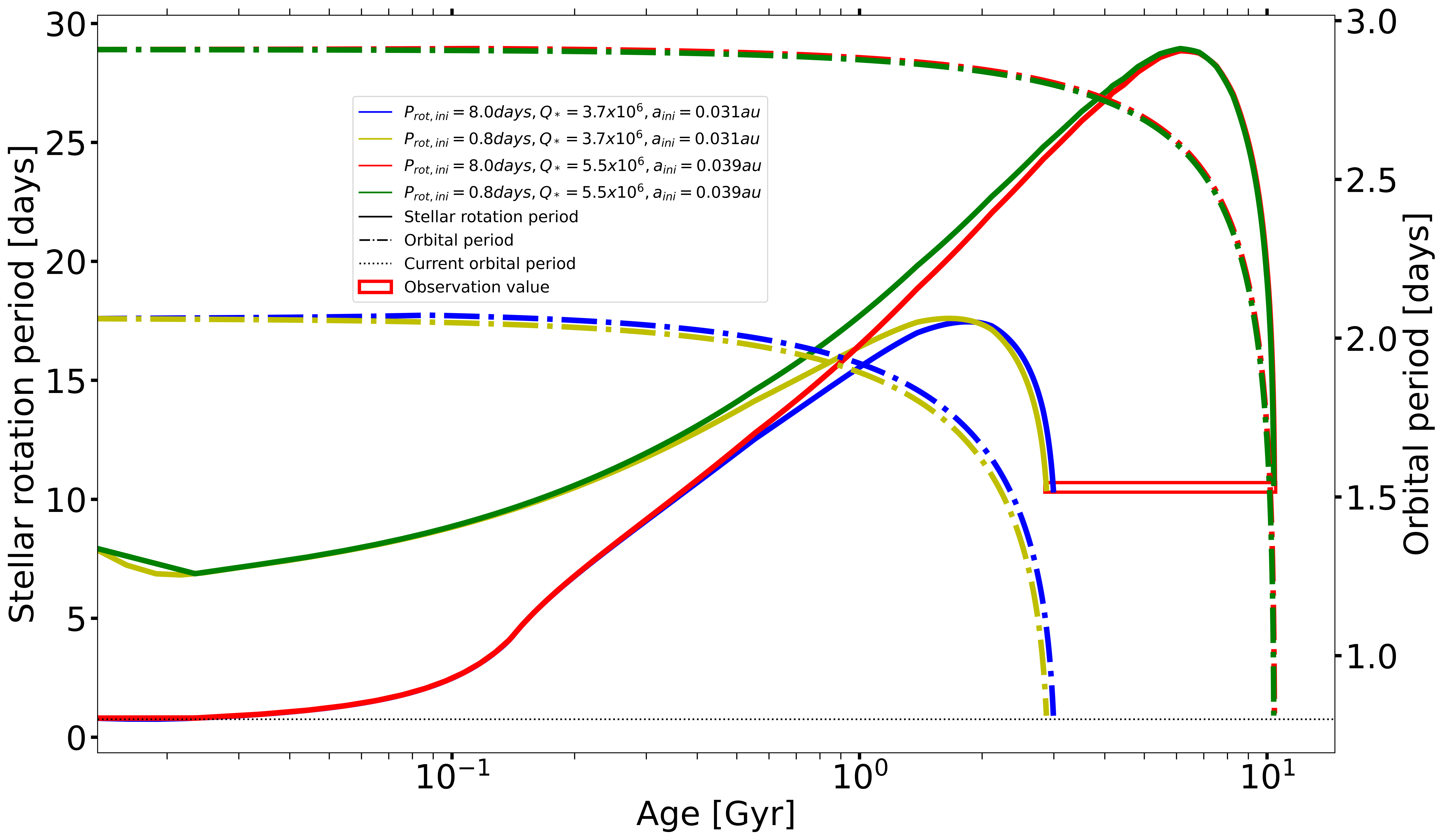}
  \end{minipage}%
	
\begin{minipage}[t]{\linewidth}
  \centering
    \caption*{(b)}
   \includegraphics[width=0.8\linewidth]{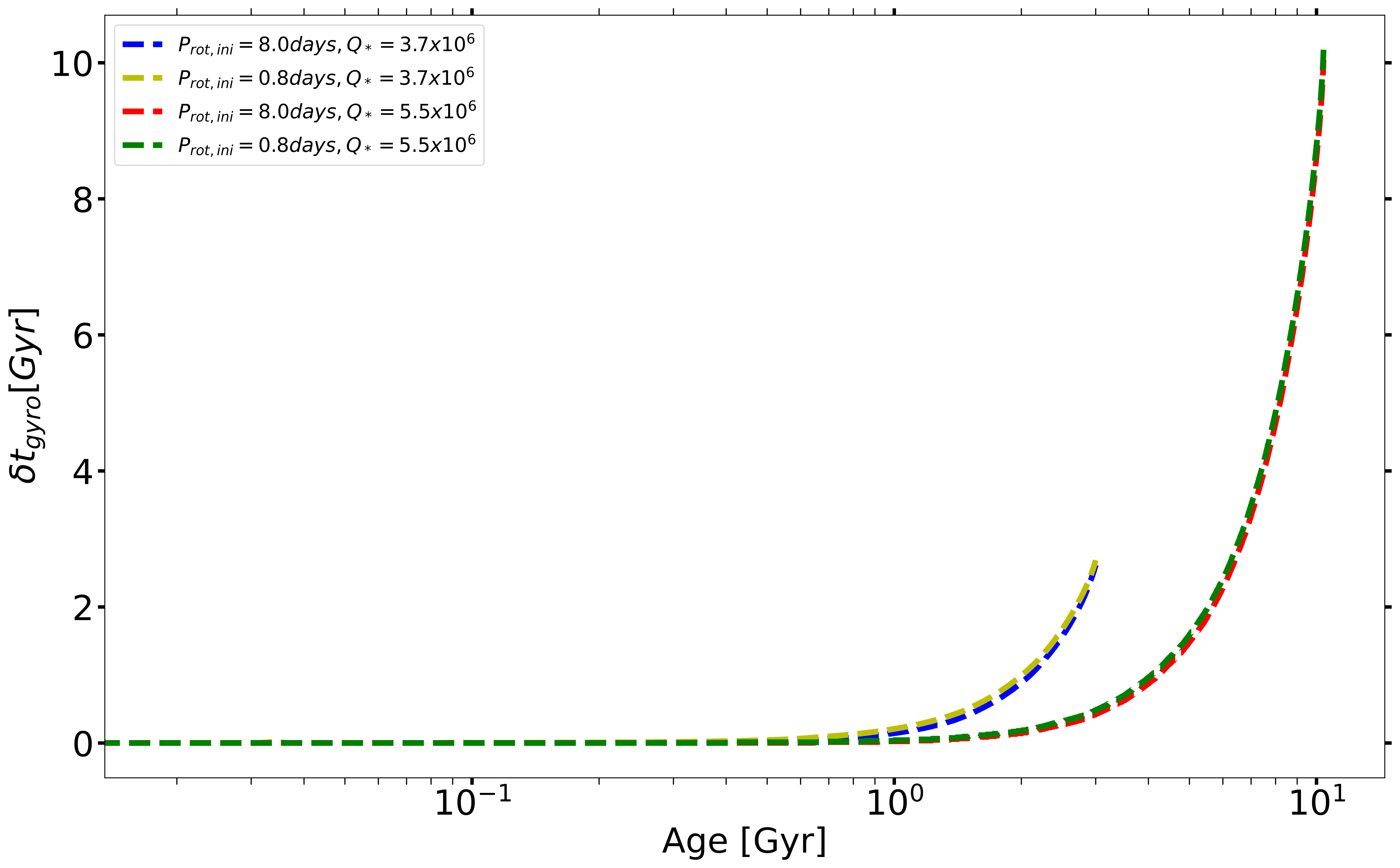}
  \end{minipage}%
    \caption{Panel (a) depicts the evolution of the planetary orbital period and stellar rotation period in the WASP-19 system. The black horizontal dotted line represents the observed value of the current orbital period, and the red rectangle represents the range of observed values for the current age and rotation period of the star. The colored dashed and solid lines represent the evolution curves of the planetary orbital period and stellar rotation period, respectively. The parameters of the red curve are ($P_{\mathrm{rot, ini}}$ = 0.8$\,$days, $Q'_{*} = 5.5 \times 10^{6}$, $a_{\mathrm{ini}}$ = 0.039$\,$au), the parameters of the green curve are ($P_{\mathrm{rot, ini}}$ = 8$\,$days,$Q'_{*} = 5.5 \times 10^{6}$, $a_{\mathrm{ini}}$ = 0.039$\,$au), the parameters of the blue curve are ($P_{\mathrm{rot, ini}}$ = 0.8$\,$days, $Q'_{*} = 3.7 \times 10^{6}$, $a_{\mathrm{ini}}$ = 0.031$\,$au), and the parameters of the yellow curve are ($P_{\mathrm{rot, ini}}$ = 8$\,$days,$Q'_{*} = 3.7 \times 10^{6}$, $a_{\mathrm{ini}}$ = 0.031$\,$au). It can be clearly seen from the figure that the differences caused by the initial stellar rotation period can be almost eliminated after 1.0$\,$Gyr, and the insensitivity of the initial stellar rotation period can more accurately predict the initial semi-major axis of the orbit and $Q'_{*}$ of this star from a theoretical perspective. Panel (b) depicts the evolution of $\delta t_{\mathrm{gyro}}$ in the WASP-19 system. Here, $\delta t_{\mathrm{gyro}} = t_{\mathrm{without}} - t_{\mathrm{with}}$, where $t_{\mathrm{without}}$ represents the age of the star obtained using the gyrochronology method without taking into account the presence of the planet, and $t_{\mathrm{with}}$ represents the age obtained with the planet. The dashed lines show the evolutionary trajectory of the star's age and $\delta t_{\mathrm{gyro}}$ under different initial parameter conditions.  \label{fig:wasp}}
   \end{figure}

From Figure \ref{fig:wasp} (a), we can see that in the late stage when the planet is close enough to the star, the angular momentum loss caused by the wind torque is much less than the angular momentum gain caused by tidal torque. And the star's rotation rate increases. Therefore, the size of the current stellar rotation period is more dependent on the tidal dissipation rate, and the observation gives a small error in the stellar rotation period of WASP-19. When the planet mass and orbital semimajor axis are fixed, the tidal quality parameter $Q'_{*}$ determines the tidal dissipation rate, which we can more easily constrain. Under our model assumptions, $Q'_{*}$ is a constant, so given a certain $Q'_{*}$, we also obtained the range of initial orbital semimajor axis. In the end, we determined that $Q'_{*} = (4.6 \pm 0.9) \times 10^{6}$ and the initial orbital semimajor axis is $(0.035 \pm 0.004)$$\,$au. The results given by \citet{2020MNRAS.491.1243P} and \citet{2022A&A...668A.114R} are  $Q'_{*} > (1.23 \pm 0.231) \times 10^{6}$ and  $Q'_{*} > (1.26 \pm 0.1) \times 10^{6}$ respectively, and they provided the minimum limit of  $Q'_{*}$ for this system in their paper. Our calculated results are within the possible range and thus have a certain degree of credibility. For the rotation-age relationship of the star, we demonstrate in Figure \ref{fig:wasp} (b) the impact of planets on the age estimation of the star. From the figure, we can see that for the simulation results of the lower age limit of observation, the age error caused by the gyrochronology method begins to appear after 1.0$\,$Gyr. For the simulation results of the upper age limit of observation, the age error caused by the gyrochronology method begins to appear after 3.0$\,$Gyr. Furthermore, as the planet orbit decays, the age error becomes larger. At the current orbital position, the age error reaches its maximum, and without considering the existence of planets, the gyrochronology method can estimate the age of the star as only a few hundred million years. In extreme cases, such as the green dashed line in the figure, where $\delta t_{\mathrm{gyro}}$ exceeds 10$\,$Gyr, the gyrochronology method can estimate the age of the star with a current age of 10.5$\,$Gyr as only 220$\,$million years.

\section{Discussion}
\label{sec:discussion}
Our results show the influence of different metallicities on stellar rotation periods and planetary engulfment timescales under star-planet interactions.

At present, the loss mechanism of stellar angular momentum has various braking models and reproduces the rotation distribution of some star clusters very well. However, with the emergence of more and more observational evidences, some current braking models still have some limitations. For example, \citet{2021ApJ...912...65G} implemented the braking model of \citet{2015ApJ...799L..23M} and \citet{2018ApJ...862...90G} in MESA for the first time and found that the rotation period of stars with an age $\geqslant$ 1.0$\,$Gyr is often overestimated. \citet{2019ApJ...879...49C} believed that at 1.0$\,$Gyr, there is a stagnation phase of angular momentum loss for stars with masses $\leq$ 1.0$\,$$M_{\odot}$ and found that gyrochronology models tend to predict too much angular momentum loss. It may underestimate the timing of the impact on the stellar rotation period after a planetary engulfment event.

Our model employs parameterized tidal quality parameter, which may not always be accurate. The advantage of using parameterized tidal dissipation factors is that we can systematically analyze the correlations between the various parameters and tidal quality parameter. Recently, some works tend to use a self-consistent modeled tidal quality parameter $Q'_{*}$ \citep{2017A&A...604A.113B,2019A&A...621A.124B,2021MNRAS.508.3408L}. However, modeled tidal quality parameter also bring a problem, as the results of \citet{2021MNRAS.508.3408L} indicate that the efficiency of dynamical tidal dissipation is stronger in metal-rich stars. This does not support the viewpoint proposed by \citet{2021MNRAS.508.3408L} that the trend of hot Jupiter occurrence rate with metallicity variation is due to the fact that metal-poor stars more frequently devour planets than metal-rich ones. \citet{2021MNRAS.508.3408L} points out in the paper that this difference can be explained by different techniques of calculating tidal quality parameter. The impact of self-consistent modeled tidal quality parameter $Q'_{*}$ goes beyond the scope of our article, and we will discuss it more deeply in our next work.

\section{Conclusions}
\label{sec:conclusion}
We implement the magnetic wind braking model of \citet{2015ApJ...799L..23M} into MESA and further incorporate star-planet tidal interaction. To do this, we focus on the main sequence stars with the spectral types of F, G and K. We calculate a large samples and focus on the effect of the tidal interaction on the stellar spin variation $\delta P$. Our simulations show that for any star-planet system with orbital decay, $\delta P$ always increases with evolution time before the planet is engulfed. We find that $\delta P$ is weakly affected by the initial rotation period of the stars for less massive stars with higher metallicities, but is greater for stars with higher masses and lower metallicities. Such behaviors are also appear in the stars with higher initial rotation periods. For a star with a smaller mass and a higher metallicity, the changes in the rotation period of the star caused by the engulf of planet will be eliminated rapidly after a planetary is engulfed. On the contrary, stars with higher masses and lower metallicities tend to produce more prominant period changes for larger initial stellar rotation period and planetary masses, and the impact of planetary engulfment lasts longer.

We also found that the convective envelope is thin enough that the effects of the initial rotation of the star and the mass of the planet are large and not negligible, and vice versa. When the planet is engulfed later, the transfer of angular momentum is mild, and the loss of angular momentum by the stellar magnetic braking is also relatively mild, and the influence of the planetary engulfment is eliminated more slowly. For stars with thin convective envelopes, the early angular momentum loss of fast initial rotation stars is severe and  the change of the rotation period of the stars is easy to be eliminated. However, the slower initial rotation of the star has weaker magnetic braking, which tends to produce larger changes in the rotation period of the star.

Finally we also simulated star-planet systems: WASP-19. For WASP-19 system, it is a G-type star with a relatively thick convection zone. And system evolution is insensitive to different initial stellar rotation periods. We estimated system's tidal quality parameter $Q'_{*} = (4.6 \pm 0.9) \times 10^{6}$ and initial orbital semi-major axis as $(0.035 \pm 0.004)$$\,$au. We also discussed how the close-in hot Jupiter affects the star in WASP-19 and how much error the gyrochronology method may produce. We found that in extreme cases, the actual age of a star over 10$\,$billion years old may be estimated as less than 220 million years, which could result in a very old star being incorrectly estimated as very young. Therefore, for systems with close-in hot Jupiters, caution is needed when using the gyrochronology method.

\appendix                  

\section{The rotation period of the Sun} \label{subsec:sun}
\begin{figure}
   \centering
   \includegraphics[width=\textwidth, angle=0]{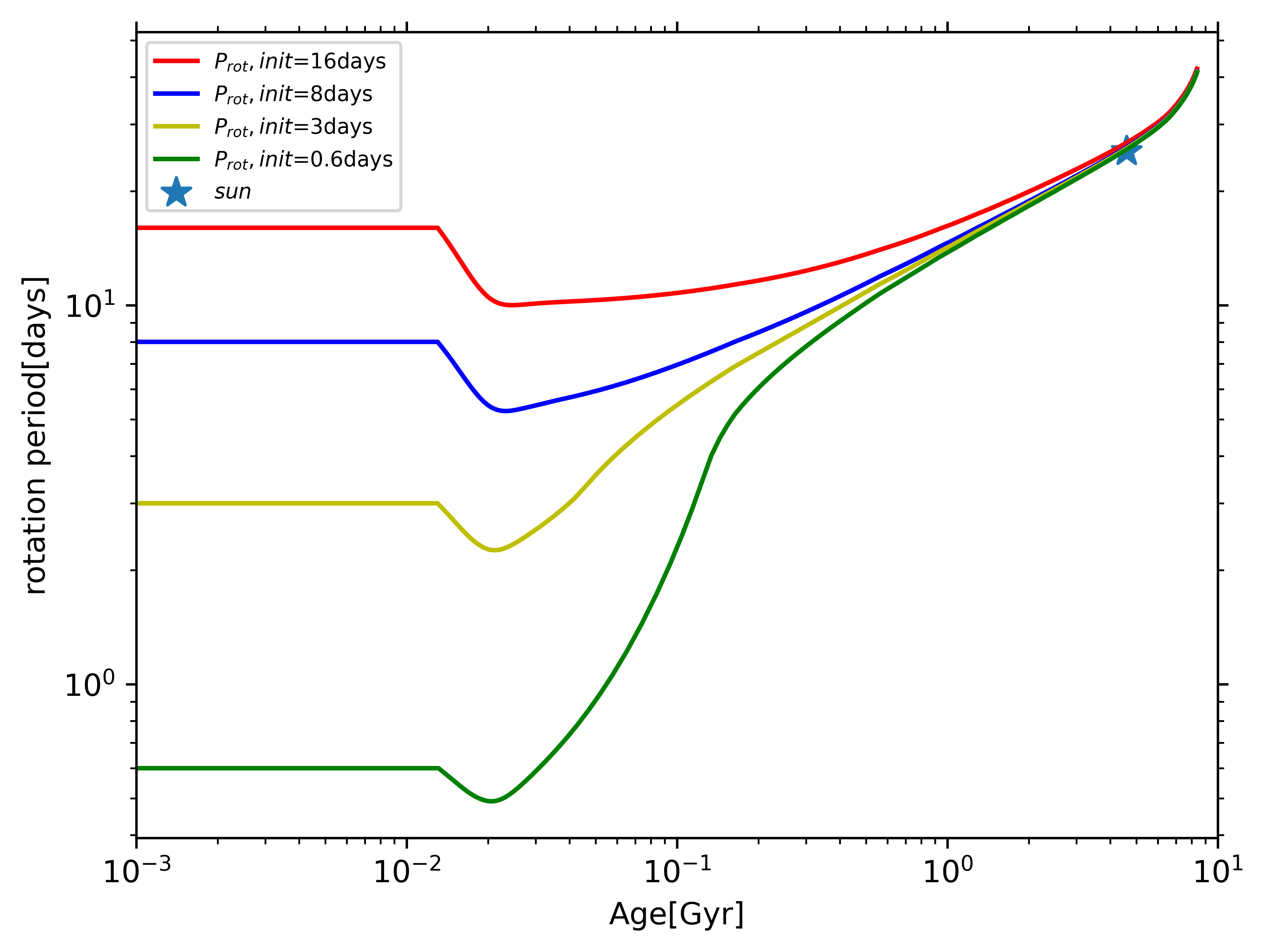}
    \caption{Evolution of the rotation period as a function of time for the model of solar mass and metallicity. The initial rotation period of each model are 0.6, 3, 8, 16$\,$days respectively. The pentagram sign represents the current position of the Sun.   \label{fig:gro}}
   \end{figure}

In this section, we first test our model by comparing it with the rotation period of the Sun. In this situation, the tide produced by the planet is ignored. Figure \ref{fig:gro} shows the rotation period as a function of evolution time for the stars with solar mass and metallicity for different initial rotation periods. It is clear that, as the stars evolve, the stellar rotation periods gradually converge to 25.4$\,$days regardless of the initial period. When the age of the star exceeds 1.0$\,$Gyr, such a behavior shows the possibility of applying gyrochronology to estimate the ages of stars if the stellar rotation periods are measured accurately. For isolated stars of F,G, K and M spectral types, the method of gyrochronology can estimate an age with an accuracy of 10 percent, which is much better than that estimated by the isochrone line with an average error of 50$\%$.  \citep{2008ApJ...687.1264M,2009MNRAS.400..451C,2011MNRAS.413.2218D}. Figure \ref{fig:gro} also shows the credibility of our model.

   .

\bibliographystyle{unsrtnat}
\bibliography{references}






\end{document}